\documentclass[9pt,twocolumn,twoside]{opticajnl}
\journal{opticajournal} 
\setboolean{shortarticle}{false}

\usepackage{lineno}
\usepackage{physics}
\usepackage[normalem]{ulem}

\newcommand{\blue}[1]{#1}

\newcommand{\purple}[1]{}
\newcommand{\teal}[1]{#1}
\newcommand{\magenta}[1]{#1}
\newcommand{\bk}[1]{\left(#1\right)}
\newcommand{\Bk}[1]{\left[#1\right]}
\newcommand{\BK}[1]{\left\{#1\right\}}

\newcommand{\MSE}{\textrm{MSE}}
\newcommand{\CRB}{\textrm{CRB}}
\newcommand{\QCRB}{\textrm{QCRB}}
\newcommand{\expect}{\mathbb E}

\newcommand{\sint}[1]{\lbrack#1\rparen}
\newcommand{\intall}{\int_{-\infty}^\infty}

\title{Passive optical superresolution at the quantum limit}

\author[1,*]{A. I. Lvovsky}
\affil[1]{Department of Physics, University of Oxford, Oxford, OX1 3PU, UK}
\affil[*]{Alex.Lvovsky@physics.ox.ac.uk}

\author[2]{Michael R. Grace}
\affil[2]{Physical Sciences and Systems, RTX BBN Technologies, Cambridge, MA 02138, USA}

\author[3]{Saikat Guha}
\affil[3]{Department of Electrical and Computer Engineering, University of Maryland, College Park MD, USA}

\author[4,5]{Mankei Tsang}
\affil[4]{Department of Electrical and Computer Engineering,
  National University of Singapore, 4 Engineering Drive 3, Singapore
  117583}
\affil[5]{Department of Physics, National University of Singapore,
  2 Science Drive 3, Singapore 117551}
  
\author[6]{Gerardo Adesso}
\affil[6]{School of Mathematical Sciences and Centre for the Mathematics and Theoretical Physics of Quantum Non-Equilibrium Systems,
University of Nottingham, University Park, Nottingham NG7 2RD, UK}

\author[7]{Nicolas Treps}
\affil[7]{Laboratoire Kastler Brossel, Sorbonne Universit\'e, CNRS,
ENS-Universit\'e PSL, Coll\`ege de France, 4 place Jussieu, F-75252, Paris, France}

\begin{abstract}
For more than a century, the diffraction limit has defined the resolution achievable by passive optical imaging systems. Although some resolution improvement can be gained through classical data processing of the image, it is limited by the noise arising from quantum nature of light. Minimizing the effect of this noise requires quantum treatment of optical imaging. By reformulating imaging as a problem of quantum measurement and estimation, it becomes possible to identify optimal detection strategies that recover spatial information previously thought inaccessible. This review summarizes the theoretical framework that underpins this development, from the formulation of quantum Cram\'er-Rao bounds and Chernoff bounds to the construction of receivers that attain them, such as those based on spatial-mode demultiplexing.  We show how these methods can beat conventional imaging in the classification, localization, and imaging of sub-Rayleigh incoherent sources. We then discuss extensions to multiparameter and partially coherent scenarios, and highlight the unifying connections between estimation and discrimination tasks. Finally, we survey recent experimental demonstrations that approach quantum-limited resolution and outline emerging applications in microscopy, astronomy, and optical sensing.
\end{abstract}

\setboolean{displaycopyright}{false} 

\begin{document}

\maketitle

\section{Introduction}
Since the invention of optical imaging devices, such as microscopes and telescopes, scientists have striven to enhance their resolution. A fundamental limitation on that resolution is imposed by diffraction on the objective lens. The diffraction limit calls for the use of higher numerical aperture objective lenses for the microscopes, and of larger objective mirrors and multi-antenna radio telescope arrays, for the telescopes. As the building of imaging devices with the required improvements became more and more expensive, the scientific community started to seek alternatives that could overcome this limit.

In the last decades, a number of techniques for circumventing the diffraction limit in microscopy have been proposed, which defined a field called {\it superresolution imaging}. These techniques can be subdivided into two large families. \emph{Near-Field} techniques such as photon tunneling microscopy \cite{guerra1990photon}, near-field scanning optical microscopy \cite{hecht2000scanning}, tip-enhanced near-field optical microscopy \cite{hartschuh2008tip}, and photon scanning tunneling microscopy \cite{ohtsu1995progress} use a probe to record the information of the evanescent waves that decay exponentially with the distance away from the sample \cite{goodman2005introduction}. As these methods use no objective lens, they are not limited by diffraction. One the other hand, \emph{far-field nonlinear microscopy}, which includes stimulated emission depletion (STED) \cite{hell1994breaking}, direct stochastic optical reconstruction (dSTORM) \cite{rust2006sub}, photoactivated localization microscopy (PALM) \cite{betzig2006imaging,hess2006ultra} and other methods enhance resolution by causing individual fluorophores to fluoresce at different moments in time and overposing the images afterwards. 

A common feature of all existing superresolution techniques is the requirement of active interaction with the specimen, either via direct access via a probe or by invoking nonlinear optical effects therein. This precludes their application in any areas where the object of investigation is inaccessible, such as remote sensing, satellite imaging or astronomy. Even in microscopy, the range of specimens that are either sufficiently robust to near-field probing or possess the required nonlinear properties is quite limited. 
The question we aim to address in this review is: can we achieve superresolution by observing the light coming from the object passively, without any active interaction with that object?

\begin{figure*}[ht]
	 \begin{center}
	 	\includegraphics[width=0.9\textwidth]{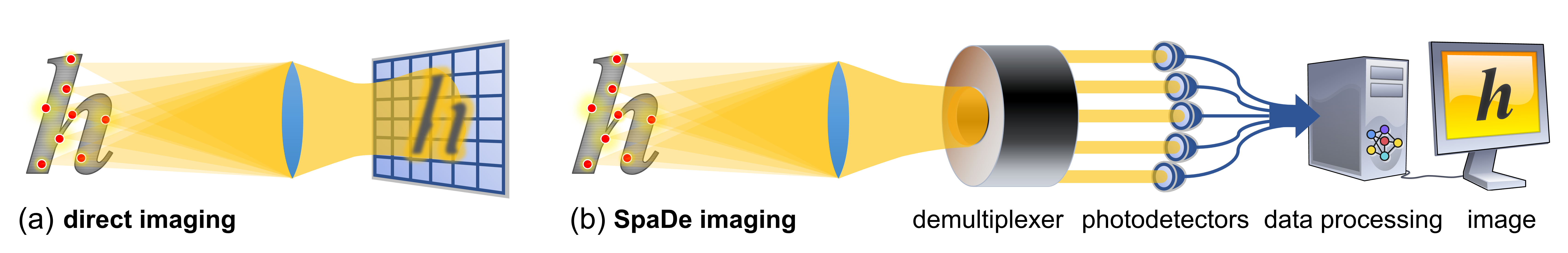}
	 	\caption{\label{fig:schemeimaging}Conceptual difference between (a) direct imaging and (b) spatial-mode demultiplexing \blue{(SPADE)}.}	     
	 \end{center}
	 \end{figure*}

Although the diffraction limit has been known for one and a half centuries years and appeared unshakeable, a recent theoretical breakthrough revealed that it can be beaten by treating imaging as a quantum sensing, i.e.,~parameter estimation problem. By transforming the state of light collected by the objective lens and subsequently measuring in a judiciously chosen basis, the amount of information about the parameters of interest can be significantly enhanced in comparison with directly detecting the light in the image plane of that lens.
Technically, this involves decomposing the incoming field into an orthonormal basis of spatial modes, typically Hermite-Gaussian (HG), and measuring the amplitude or intensity of each basis component (Fig.~\ref{fig:schemeimaging}). From these measurements, parameters of the original object can be reconstructed. This method of \emph{spatial demultiplexing}, or \emph{SPADE}, enables us to not only achieve sub-Rayleigh precision, but also, in some cases, reach the ultimate resolution limits allowed by quantum mechanics. 

The key intuition behind this enhanced resolution is that the optimum modal basis in which to detect the information bearing light is a strong function of the quantitative task at hand (e.g., estimation of a scene parameter, classification, localization of a known feature, counting, etc.) and that the (inevitable) detection shot noise irreversibly corrupts information going from the optical to the electronic domain. Therefore photon detection in the task-specific information-optimal basis is key, as the resolution-enhancing effect of spatial-mode transformation cannot be replicated in the electronic domain once shot noise---from detection in a sub-optimal basis---has irreversibly corrupted the information in the light collected by the imager.

\section{Ultimate limits to optical sensing}\label{sec:O1}

Traditional resolution criteria --- such as those introduced by Abbe \cite{Abbe1873}, Rayleigh \cite{Rayleigh79}, and Sparrow \cite{Sparrow1916} --- are related to the width of the point-spread function (PSF) of the imaging system. The PSF width is inversely proportional to the bandwidth $\Delta k$ of the optical transfer function (OTF), which is the Fourier transform of the PSF.  Physically, the bandwidth is limited by the numerical aperture
\cite{goodman2005introduction}. These criteria, though
influential, apply specifically to direct imaging, and are inadequate if sophisticated signal pre- or post-processing is allowed. For example, 
if the image is acquired with suppressed noise, digital post-processing can enhance the fundamental resolution through deconvolution techniques \cite{villiers}. Lastly, and crucially for the purposes of this review, in many cases the goal is not to produce a high resolution image but to infer some quantitative properties about the scene, in which case error metrics from the fields of estimation and detection theory will more faithfully quantify the sensing performance than traditional resolution criteria. Hence, to define the ultimate resolution, one should
account for not only the PSF but also the statistical model of the
analog measurement, the digital data processing method, and the quantitative imaging problem at hand.

\subsection{Quantum estimation and detection theory tutorial}
We begin by defining a few basics of the quantum sensing theory: the field that studies statistical decision problems, whose goal is to learn unknown features of the object using observations affected by quantum noise. Two classes of problems can be defined: \emph{estimation}, aimed at evaluating continuous parameters of the object, and \emph{hypothesis testing}, aimed at choosing the most likely hypothesis amongst a discrete set. For example, observing a point-like object in the night sky, we can test the hypothesis whether the object is a single star or a binary one (hypothesis testing) or, given the prior information that the object is binary, estimate the distance between its components (estimation).


\subsubsection{Estimation}\label{sec:est}
Continuous parameter estimation errors can be evaluated
using standard statistical benchmarks, such as the \emph{Cram\'er-Rao
  bound (CRB)}, which gives a lower limit to the mean-square error
of that estimation. The CRB has become one of the most popular tools
for defining resolution in modern imaging research
\cite{zmuidzinas03,feigelson,chao16,diezmann17}. In single-molecule
microscopy, in particular, the bound is the standard proxy and
benchmark for parameter-estimation imprecision
\cite{chao16,diezmann17}.

The CRB is defined as follows. \textcolor{black}{Consider  a physical system, which  depends on a set of unknown continuous parameters $\boldsymbol{\theta}=(\theta_1,\dots,\theta_n)$. Suppose a set of $N$ independent, identical measurements on that system produced results $\mathbf{y}=(y_1,y_2,\dots,y_N)$. Knowing the statistical model of the measurement, i.e.~the probability distribution $p(y|\boldsymbol{\theta})$ of obtaining a  measurement result $y$ conditioned on a particular set $\boldsymbol{\theta}$ of parameter values, one wishes to learn the parameters from the observations.}

\teal{To this end, 
one computes an \emph{estimator} $\check{\boldsymbol{\theta}}(\mathbf{y})$ --- a guess of the parameter set based on the observations. The mean-square error of the estimator, 
\begin{align}
\MSE_j(\boldsymbol \theta) &\equiv \expect_{\boldsymbol\theta}\BK{\Bk{\check{\theta_j}(y_1,\dots,y_N)-\theta_j}^2},
\end{align}
consists of the diagonal entries of the covariance matrix $\text{Cov}[\check{\boldsymbol{\theta}}]=\expect_{\boldsymbol{\theta}}\left\{(\check{\boldsymbol{\theta}}-\boldsymbol{\theta})\otimes(\check{\boldsymbol{\theta}}-\boldsymbol{\theta})\right\}$,
where $\expect_{\boldsymbol{\theta}}$ denotes the expectation value with respect to $p(y|\boldsymbol{\theta})$ and $\otimes$ outer product. 
For any unbiased estimator ($\expect(\check{\boldsymbol{\theta}}) = \boldsymbol{\theta}$),
the CRB is given by
\begin{align}
\text{Cov}[\check{\boldsymbol{\theta}}] \ge \frac1N  I(\boldsymbol{\theta})^{-1}
\equiv \CRB(\boldsymbol{\theta}),
\label{CRB}
\end{align}
where 
\begin{align}\label{CFI}
    I_{ij}(\boldsymbol{\theta})\equiv \expect_{\boldsymbol{\theta}}
\Bk{\pdv{\ln p(y|\boldsymbol{\theta})}{\theta_i}\pdv{\ln p(y|\boldsymbol{\theta})}{\theta_j}}
\end{align}
is} the \emph{Fisher information} (FI) matrix $I(\boldsymbol{\theta})$. Notice that, apart from the
unbiased assumption, the CRB does not depend on the estimator
$\check\theta$, so it offers a convenient proxy of the error given a
measurement model without the need to account for the data processing
method. 

It can be shown that, under rather general conditions, the error of
the maximum-likelihood estimator can asymptotically achieve the bound
in the $N\to\infty$ limit \cite{LehmannCasella}, so the CRB is not
only a lower error bound but also the asymptotic error of a standard
estimator. Note also that the Fisher information can be used to
determine Bayesian and minimax lower error bounds that forgo the
unbiased-estimator assumption
\cite{VanTrees2007bayesian,tsang18a,lee2022quantum}.

This review covers a practical case where the parameter estimation precision is limited by
quantum noise. When the optical energy arriving from the scene and captured by the imaging system (e.g., telescope or microscope) is finite, any measurement will be contaminated by the
shot noise that arises from the discrete and random nature of
photon detection. This noise inevitably translates into an estimation
error. In a quantum mechanical treatment, the outcome distribution $p(y|\boldsymbol{\theta}) \equiv {\rm Tr}\left(\hat\rho(\boldsymbol{\theta})\Pi_y\right)$ is a function of the quantum state of light before the measurement $\hat\rho(\theta)$ and a positive operator valued measure (POVM) $\left\{\hat{\Pi}_y\right\}$ associated with the physical measurement. The task confronting an experiment designer is not only to find the estimator that minimizes the uncertainty of estimating $\boldsymbol{\theta}$ for a particular measurement with known $p(y|\boldsymbol{\theta})$, but also 
to find the measurement that minimizes the uncertainty given the quantum optical state of the light obtained from the imaging system.

The latter task is daunting because, in principle, there are
infinitely many optical components, linear or nonlinear, that one can
pick or conceive, as well as infinitely many ways to combine them to
construct a measurement device. Fortunately, the quantum detection and
estimation theory pioneered by Helstrom offers an elegant approach
\cite{Helstrom1976}, playing a fundamental role in sensing and imaging
not unlike the role of thermodynamics in engine design.  Given the
quantum state, Helstrom's theory enables one to compute lower bounds
on the decision errors for \emph{any} measurement allowed by quantum
mechanics. 

\textcolor{black}{Specifically, Helstrom constructed the \emph{quantum Cram\'er-Rao
  bound (QCRB)} \cite{Helstrom1976,Helstrom1976,nagaoka89,BraunsteinCaves,hayashi}.} 
  Akin to its
classical counterpart, the QCRB depends on the inverse of the
\emph{quantum Fisher information} (QFI) matrix. To determine the QFI
matrix, one considers small variations $\Delta\boldsymbol{\theta}$ of the
parameters and evaluates the effect of this change on the state
$\hat\rho(\boldsymbol{\theta})$. The effect can be quantified by the fidelity
$F(\hat\rho(\boldsymbol{\theta}),\hat\rho(\boldsymbol{\theta}+\Delta\boldsymbol{\theta}))$, defined as
$F(\hat\rho_1,\hat\rho_2)\equiv\bk{{\rm
Tr}\sqrt{\sqrt{\hat\rho_1}\hat\rho_2\sqrt{\hat\rho_1}}}^2$. Intuitively, the
fidelity measures the indistinguishability of two states: it is equal
to $1$ for $\Delta\boldsymbol{\theta}=0$ and decreases quadratically for small
$\Delta\boldsymbol{\theta}$. The QFI matrix, defined as
\begin{align}\label{IQ}
    I_{\rm Q}(\boldsymbol{\theta})_{ij}=-2\left.
\pdv[2]{F(\hat\rho(\boldsymbol{\theta}),
\hat\rho(\boldsymbol{\theta}+\Delta\boldsymbol{\theta}))}{\Delta\boldsymbol{\theta}_i}{\Delta\boldsymbol{\theta}_j}
\right|_{\Delta\boldsymbol{\theta}=0},
\end{align}
gives the coefficients for this quadratic decrease,
therefore capturing the sensitivity of the quantum state to the parameters. It can be proved that, for any physical   measurement, $I_{\rm Q}(\boldsymbol{\theta})-I(\boldsymbol{\theta})$ is a positive-semidefinite matrix. The CRB associated with any particular measurement on $\hat\rho(\boldsymbol{\theta})$ therefore obeys~\cite{nagaoka89,BraunsteinCaves,hayashi}
\begin{equation}
    \CRB(\boldsymbol{\theta})\ge\QCRB(\boldsymbol{\theta}),
    \label{eq:QCRB_bound}
\end{equation}
where
\begin{align}
\QCRB(\boldsymbol{\theta}) \equiv  
\frac{1}{N}   I_{\rm Q}(\boldsymbol{\theta})^{-1}.
\end{align}
\teal{When $\boldsymbol{\theta}$ is a single scalar parameter, then a projective measurement described by the eignevectors of the so-called {\em symmetric logarithmic derivative} (SLD) of $\rho(\boldsymbol{\theta})$, applied on each of the $N$ copies of $\rho(\boldsymbol{\theta})$, achieves QCRB$(\boldsymbol{\theta})$~\cite{Helstrom1976}.
However, this measurement may generally need to be adaptive, i.e.~with the apparatus being modified as results are acquired based on the analysis of these results, because the SLD eigenbasis in general depends on the parameter $\boldsymbol{\theta}$ itself. }

\teal{When $\boldsymbol{\theta}$ is a multi-parameter set, the SLDs for the parameters may not commute with one another. In that case, in general, even a collective measurement acting on all $N$ copies of $\rho(\boldsymbol{\theta})$ may not achieve CRB $=$ QCRB. However, if the SLD operators $L_{\theta_i}$ and $L_{\theta_j}$ for some pair of parameters $(\theta_i,\theta_j)\in\boldsymbol{\theta}$ satisfy ${\rm Tr}\left(\rho(\boldsymbol{\theta})[L_{\theta_i},L_{\theta_j}]\right) = 0$ (condition known as {\em weak commutativity}), there exists a {\em joint} measurement acting on all $N$ copies of $\rho(\boldsymbol{\theta})$ that can simultaneously attain the QCRB for both parameters.}

In practice, an experiment designer can be assured that, if a measurement whose CRB equals the QCRB is found, no further fundamental sensitivity improvements can be made to the measurement apparatus for that imaging task. Conversely, a gap between the QCRB and the CRB of the state of the art means that there may be room for
further technological development to improve the imaging sensitivity.


\subsubsection{Hypothesis testing}\label{sec:class}
\teal{Unlike estimation, where the state of the system depends on a set of continuous parameters, classification (hypothesis testing) considers systems that can be in a finite set of $M$ discrete states. Again, suppose we perform $N$ identical measurements, obtaining results $\mathbf{y}=(y_1,y_2,\dots,y_N)$, which are all drawn from one of $M$  distributions $p_j(y)$, with $1\le j \le M$. Our goal is to pick the hypothesis (the value of $j$) with the minimum error.} We consider the average probability of error $P_e$, which arises for Bayes tests in symmetric detection scenarios~\cite{VanTrees2013}. In the limit $N\to\infty$, this probability has been proven to decay exponentially with $N$ as
\begin{equation}
    P_{\rm e} = \exp[-\xi N+o(N)],
    \label{eq:ChernoffBound}
\end{equation}
where the decay rate is set by the {\em Chernoff Exponent} (CE) $\xi$. For the binary decision case ($M=2$), the CE is given by~\cite{VanTrees2013} 
\begin{equation}
    \xi\equiv -\log\left[\min_{0\leq s \leq 1}\left\{\int p_1(y)^s p_2(y)^{1-s}dy\right\}\right].
    \label{eq:CE}
\end{equation}

Like with estimation, quantum detection theory can be used to bound the probability of error over all possible measurements of the collected light. To evaluate this limit, the hypotheses are cast as a prior-known library of candidate quantum states ${\hat \rho}_j^{\otimes N}$ consisting of $N$ identical objects each in the state ${\hat \rho}_j$, with $1\le j \le M$. 
The CE of any particular measurement obeys the \emph{quantum Chernoff bound}
~\cite{kargin2005chernoff,nussbaum_chernoff_2009,audenaert2007}
\begin{equation}
    \xi\leq\xi_{\rm Q},
    \label{eq:QCB}
\end{equation} 
where the \emph{quantum Chernoff exponent} (QCE) is given by~\cite{audenaert2007}
\begin{equation}
\xi_{\rm Q} \equiv -\log \left[\min_{0 \le s \le 1}{\rm Tr}\left({\hat \rho}_1^s{\hat \rho}_2^{1-s}\right)\right]
\end{equation}
in the $M=2$ binary case. When one of the two states is a pure state, say $\hat{\rho}_1=\ket{\psi}\bra{\psi}$, the QCE can be directly related to the fidelity $F(\hat{\rho}_1,\hat{\rho}_2)=\langle{\psi}\vert{\hat{\rho}_2}\vert{\psi}\rangle$ via the simpler expression~\cite{kargin2005chernoff}
\begin{equation}
    \xi_{\rm Q}=-\log\left(\langle{\psi}\vert{\hat{\rho}_2}\vert{\psi}\rangle\right).
    \label{eq:QCB_pure}
\end{equation}
The QCE plays a role as an information metric much like the QFI, quantifying the maximum efficiency with which each additional quantum object drives down the average error. In other words, the quantum Chernoff bound serves as a hypothesis testing equivalent of~\eqref{eq:QCRB_bound}, quantifying the quantum-optimal error decay rate and precluding the existence of measurements with superior asymptotic error.

\subsection{Estimation in diffraction-limited optics}
\subsubsection{Point Source Localization}
As an example of an estimation problem, consider determining the one-dimensional coordinate $x$ of a point \teal{thermal} source imaged by a far-field objective lens. Assume the paraxial
approximation and suppose that the average photon number per optical
mode received by the imaging system is much less than one, a typical
scenario in both optical astronomy \cite{Goodman_book,zmuidzinas03}
and fluorescence microscopy \cite{chao16,diezmann17} called the
Poisson limit. \teal{In this regime, the thermal state emitted by the source into each temporal mode can be considered as an incoherent mixture of vacuum and single-photon states. Because the vacuum state carries no spatial information, the number $N$ of copies of the input optical state available for measurement is equal to the number of photons entering the objective lens of the imaging device} 
\cite{tsang11,Tsang16}.  

For the sake of simplicity, assume that the
object and image planes are one-dimensional. Our task is to estimate
the one-dimensional position $x_0$ of the
source. To compute the QCRB, assume a real PSF $\psi(x)$ for the
field, such that $\psi(x-x_0)$
is the transverse wavefunction of each photon on the image plane.
Define also
\begin{align}\label{Deltak}
\Delta k &\equiv \Bk{\intall \abs{\pdv{\psi(x)}{x}}^2 \dd{x} }^{1/2}
\end{align}
as the root-mean-square spatial bandwidth of the PSF.
The fidelity between two pure states takes the simple form
\begin{align}
F=\abs{\int_{-\infty}^\infty \psi^*(x-x_0)\psi(x-(x_0+\Delta x_0))
\dd{x}}^2\approx 1 - \Delta k^2 \Delta x_0^2,
\end{align}
from which we find via \eqref{IQ} that
\begin{equation}\label{locQFI}
I_{\rm Q}(x_0) = 4\Delta k^2,
\quad
\textrm{QCRB}(x_0) = \frac{1}{4 N \Delta k^2},
\end{equation}
where $N$ is the total number of photons arriving on the image
plane. 

The quantum bound can be achieved by direct imaging
\cite{helstrom70,chao16}. To see this, consider a camera on the image
plane detecting arriving photons. The probability density of the
position $x$ of each detected photon is given by
$p(x|x_0)=\abs{\psi(x-x_0)}^2$. Substituting this into
\eqref{CFI}, we find $I(x_0)=4\Delta k^2$. In other words, direct
detection in the image plane is \emph{quantum-optimal} for the task of \emph{localizing}
a single point source \blue{(assuming idealized conditions of absent technical noise and infinite detector resolution)}. A similar conclusion can be reached for a
thermal state after taking the Poisson limit,
where $N$ is the average photon number
received in all modes \cite{helstrom70,Helstrom1976}.

This example illustrates our earlier point regarding the insufficiency of traditional
resolution criteria, which do not scale with the photon number. The $O(1/N)$
localization error is in fact the basis of localization microscopy
\cite{betzig2006imaging,hess2006ultra,rust2006sub}, where one enforces
only a sparse subset of fluorescent particles to emit for each image
and the accuracy of locating sparse point sources can be enhanced by
collecting more photons.

\subsubsection{Two Point Source Separation Estimation}\label{sec:psse}
\begin{figure}[htbp!]
\centerline{\includegraphics[width=0.48\textwidth]{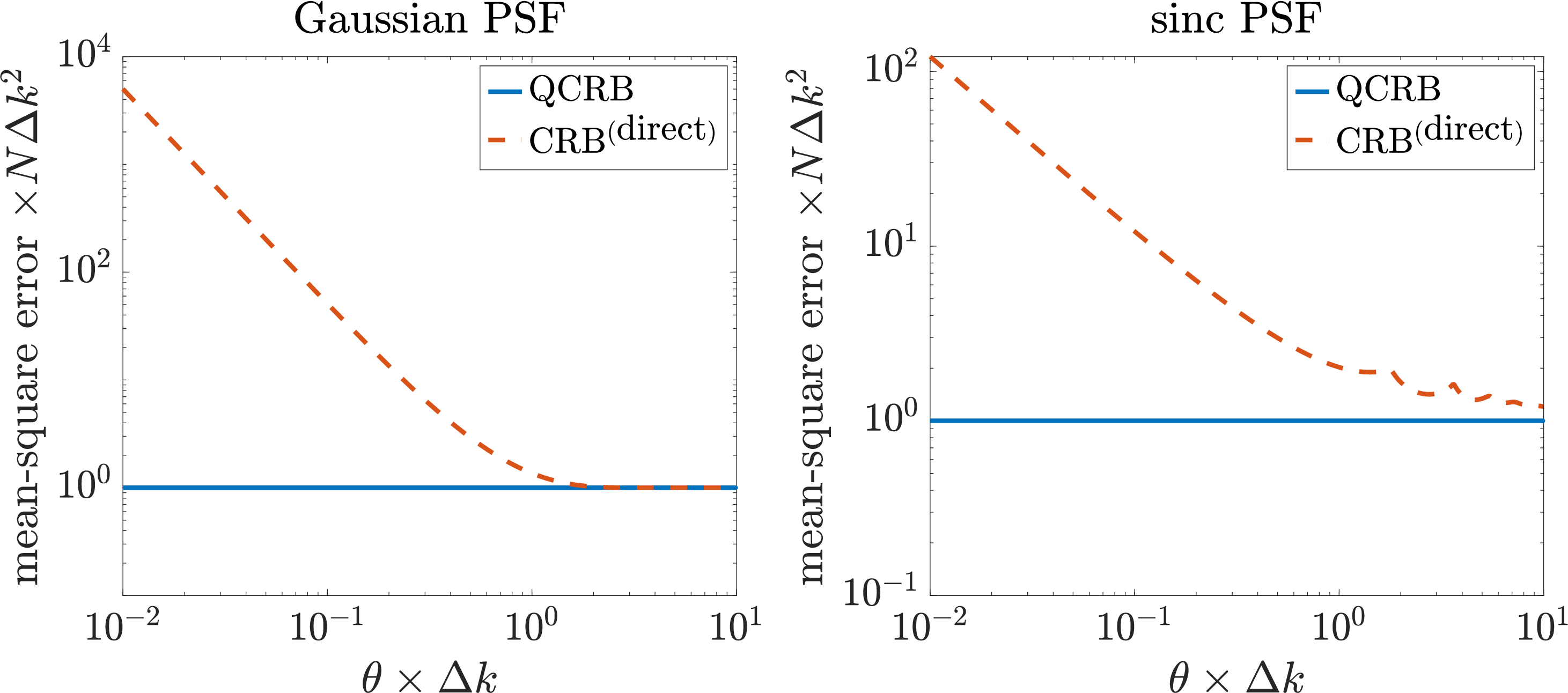}}
\caption{\label{QCRB}Log-log plots of the Cram\'er-Rao bound for
  direct imaging ($\textrm{CRB}^{(\textrm{direct})}$) and the quantum
  Cram\'er-Rao bound (QCRB) on the mean-square error of estimating the
  two-point separation $\theta$.  The axes are normalized with respect
  to the root-mean-square bandwidth $\Delta k$ of the optical transfer
  function (OTF), while the vertical axis is further normalized with
  respect to the average photon number $N$ received in all modes.  The
  left figure assumes a Gaussian PSF or
  equivalently a Gaussian OTF, while the right figure assumes a sinc
  PSF or equivalently a rectangular OTF.}
\end{figure}

Direct imaging is not quantum-optimal in all cases, as evidenced by
the following example.  Consider the case of two incoherent point
sources, as originally conceived by Lord Rayleigh
\cite{Rayleigh79}. According to Rayleigh's criterion, the scene is no longer "resolved" when the separation distance $\theta$ between the two point sources is less than a heuristic distance relative to the width of the PSF, which corresponds to the condition $\theta \Delta k<1.81$ (or, equivalently, $\theta<1.22\lambda/D$ for an angular separation $\theta$ and a hard circular aperture of diameter $D$ collecting light at wavelength $\lambda$). As we have seen, the framework of (quantum) estimation theory gives more precise criteria for resolvability in terms of achievable estimator error. Assume, for simplicity, that the two sources are
equally bright, their centroid is at $x=0$, and the separation between
the two sources $\theta \in \sint{0,\infty}$ is the only
unknown parameter to be estimated. Now the probability density of each
photon position measured by direct imaging equals
\begin{align}
p(x|\theta)=\frac12 \Bk{\abs{\psi\bk{x-\frac{\theta}{2}} }^2+
\abs{\psi\bk{x+\frac{\theta}{2}} }^2}.
\label{p_direct}
\end{align}
The Fisher information, calculated via \eqref{CFI}, yields
$I_{\rm Direct}(\theta)=\Delta k^2$ for $\theta\gg 1/\Delta k$, but \teal{(assuming that the PSF has no zeros \cite{Paur2019})} tends to zero according to $I_{\rm Direct}(\theta)=(\theta\Delta k)^2/2$ for
small source separations $\theta\to0$.  In other words, the
Cram\'er-Rao bound blows up for sub-Rayleigh separations --- a
behavior dubbed by Tsang and coworkers as \emph{Rayleigh's curse}
\cite{Tsang16} --- as shown in Fig.~\ref{QCRB}. This bound
has been proposed as a basis for an alternative fundamental resolution
measure to supersede Rayleigh's criterion
\cite{tsai79,bettens,vanaert,ram}.

Let us now compute the QCRB for this setting. The quantum state of each collected temporal mode that contains a single photon, expressed in the image plane, is given by~\cite{Tsang16} 
\begin{align}
\hat\rho(\theta)=\frac12 
\bk{\ket{\psi_{\theta/2}}\bra{\psi_{\theta/2}}+\ket{\psi_{-\theta/2}}\bra{\psi_{-\theta/2}}},
\label{eq:singlephotonstate}
\end{align}
where $\ket{\psi_{\pm\theta/2}}$ is the state of a single photon in the PSF mode $\psi(x\pm\theta/2)$, corresponding to the photon coming from the source at $\pm\theta/2$. 
The state is a mixture of the two possibilities because the sources are incoherent. Applying Uhlmann's theorem \cite{uhlmann}, we find
that the fidelity is given by
\begin{align}
F \approx 1 - \frac{1}{4} \Delta k^2\Delta\theta^2,
\end{align}
leading to 
\begin{align}
I_{\rm Q}(\theta) &= \Delta k^2,
&
\textrm{QCRB}(\theta) &= \frac{1}{N\Delta k^2}.
\label{QCRB_sep}
\end{align}
Remarkably, the \emph{quantum} bound for this setting remains constant
for any separation and shows no sign of Rayleigh's curse. Tsang, Nair,
and Lu \cite{Tsang16} first obtained this bound by assuming
the Poisson limit for the quantum state of thermal light
\cite{tsang11,tsang21}; it has been confirmed and generalized by
calculations that forgo the Poisson approximation
\cite{Nair16,Lupo16}. This result means that one can substantially
improve the estimation of the two-point separation if a
quantum-optimal measurement can be found.

Tsang \emph{et al.}\ further discovered that demultiplexing the
image-plane photons in terms of a judicious spatial-mode basis,
followed by photon counting for each mode, can achieve the quantum
limit; they called this measurement method Spatial-mode Demultiplexing, or SPADE (see Fig.~\ref{fig:schemeimaging}). For a Gaussian PSF 
\begin{equation}\label{GaussPSF}
\teal{    \psi(x)=(2\pi\sigma^2)^{-1/4}e^{-\frac{x^2}{4\sigma^2},}}
\end{equation}
\teal{where $\sigma=1/2\Delta k$,} an optimal spatial-mode basis is the set of Hermite-Gaussian modes, with
the fundamental mode matching $\psi(x)$. The resulting CRB
saturates the quantum bound \eqref{QCRB_sep} for all separations $\theta$ {(i.e., $I_{\rm SPADE}(\theta)=I_{\rm Q}(\theta)$)}, therefore eliminating Rayleigh's curse. \teal{This result was later generalized to a wide class of PSFs, including} the sinc PSF 
arising from a rectangular 1D aperture~\cite{Rehacek17.1,Kerviche2017a}. 

\begin{figure}[htbp!]
\centerline{\includegraphics[width=0.48\textwidth]{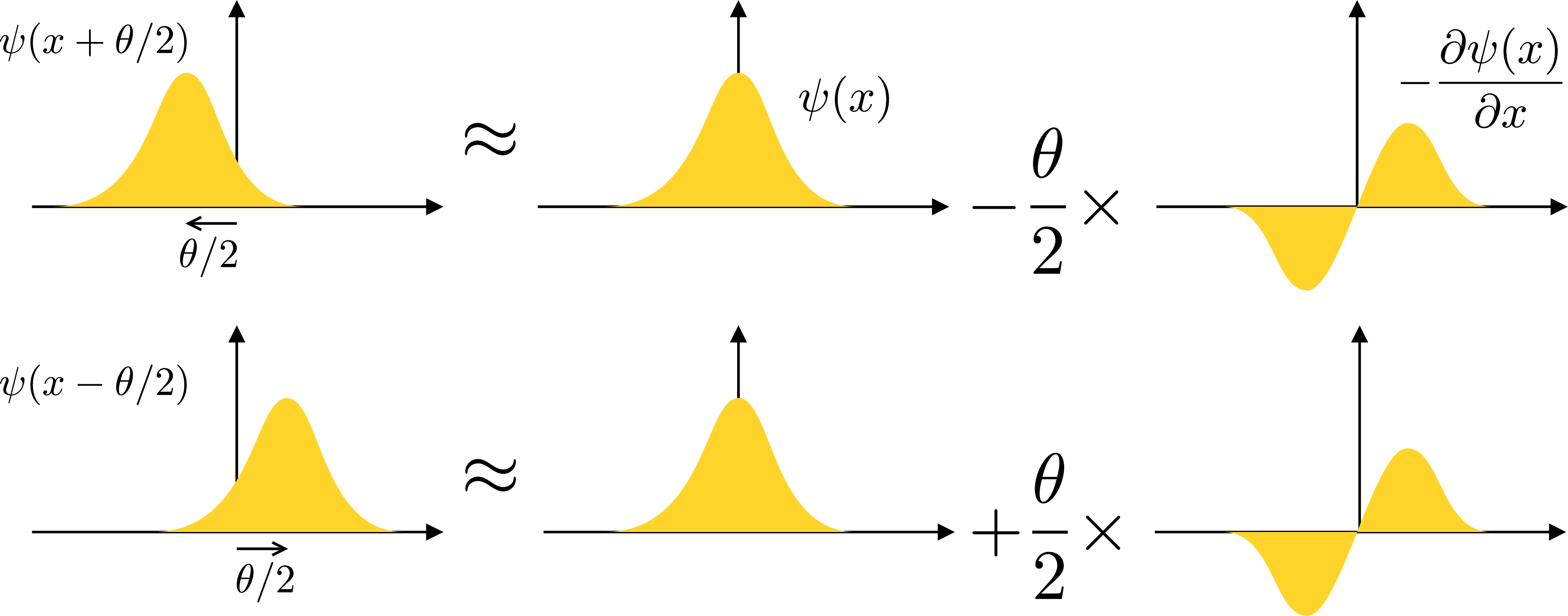}}
\caption{\label{spade_explain1}The optical field due to each point
  source can be decomposed in terms of the fundamental mode with mode
  function $\psi(x)$ and the derivative mode with mode function
  $\psi_1(x) \propto -\partial\psi(x)/\partial x$ for a sub-Rayleigh
  separation $\theta\ll 1/\Delta k$. With incoherent sources, the
  total energy in the derivative mode consists of the incoherent
  contributions from the sources and remains sensitive to the
  separation $\theta$, while the fundamental mode acts as background
  noise.}
\end{figure}

To obtain some intuition about why SPADE is optimal, let us assume a
sub-Rayleigh separation $\theta \ll 1/\Delta k$ and make the Taylor
approximation
\begin{align}\label{Taylor}
\psi\bk{x\mp\frac{\theta}{2}} &\approx \psi(x) 
\mp \frac{\theta}{2} \pdv{\psi(x)}{x}
\end{align}
for the photon wavefunction due to each source. Only two spatial modes
are excited significantly: the fundamental mode with mode function
$\psi(x)$ and the derivative mode with mode function
$\psi_1(x) \equiv -(1/\Delta k) \pdv*{\psi(x)}{x}$.  Assume that the
two modes are orthogonal ($\intall \psi_1^*(x) \psi(x) \dd{x} = 0$),
which is true if $\psi(x)$ is even for example. Eq.~(\ref{Taylor})
implies that, to the first order, the amplitude of the fundamental
mode is insensitive to $\theta$, while the amplitude of the derivative
mode due to each source is proportional to $\theta$, as
  illustrated by Fig.~\ref{spade_explain1}. For
  each of the two sources taken individually, the probability of
detecting the photon in the derivative mode is
\begin{align}\label{p10}
\abs{\intall \psi_1^*(x) \psi\bk{x \mp \frac{\theta}{2}} \dd{x}}^2
\approx \frac{\Delta k^2\theta^2}{4}.
\end{align}
Since the state is a mixture of the two possibilities that the photon
may come from either source, we average the two probabilities to find
that the unconditional probability of detecting the photon in the derivative mode remains
\begin{align}
p(1|\theta) &= \frac{1}{2} \left[\abs{\intall \psi_1^*(x) \psi\bk{x-\frac{\theta}{2}}
 \dd{x}}^2 \right.
\nonumber\\
&\quad + 
\left.\abs{\intall \psi_1^*(x) \psi\bk{x+\frac{\theta}{2}}
\dd{x}}^2\right] \approx \frac{\Delta k^2\theta^2}{4}.
\label{p1}
\end{align}
Hence, the parameter can be estimated by counting the number of
photons in the derivative mode, and the Fisher information due
to the mode becomes
\begin{align}\label{derivmode}
I(\theta) &\approx
\frac{1}{p(1|\theta)} \Bk{\pdv{p(1|\theta)}{\theta}}^2 \approx \Delta k^2,
\end{align}
which coincides with the QFI given by Eq.~(\ref{QCRB_sep}). The
information due to all other modes turns out to be negligible in the
sub-Rayleigh regime.

Contrast Eq.~(\ref{p1}) with the probability density $p(x|\theta)$ in
Eq.~(\ref{p_direct}) for direct imaging, which can be approximated as %
\begin{align}
p(x|\theta) &\approx \abs{\psi(x)}^2 +\frac{\theta^2}{8} \pdv[2]{}{x}
\abs{\psi(x)}^2.
\end{align}
This expression contains a parameter-insensitive background $|\psi(x)|^2$ due to
the fundamental mode, leading to vanishing Fisher information for
$\theta \to 0$ and thus Rayleigh's curse. With photon counting in the
derivative mode, on the other hand, any background due to the
fundamental mode is removed from Eq.~(\ref{p1}), thanks to the
orthogonality between $\psi_1(x)$ and $\psi(x)$.

For higher separations, higher-order terms in the Taylor series (\ref{Taylor}) become significant, leading to significant excitations of higher-order spatial modes. 
Generalizing the findings above, Rehacek \emph{et al.} and Kerviche
\emph{et al.} proposed general conditions on the mode basis to
saturate the QCRB for any symmetric PSF~\cite{Rehacek17.1,Kerviche2017a}.

\subsection{1-vs-2 point source discrimination}\label{sec:1vs2discrimination}

The most fundamental hypothesis test for evaluating imaging resolution is the problem that spurred Rayleigh's famous resolution criterion: one-vs-two point source discrimination~\cite{Rayleigh79}. The quantum limit of this problem was first evaluated in the concurrent papers of Krovi {\em et al.}~\cite{Krovi2016} (using the full thermal-state model) and Lu {\em et al.}~\cite{Lu2016} (using the single-photon approximation as in~\eqref{eq:singlephotonstate}), whose results were combined in Ref.~\cite{lu2018quantum}.

Let hypothesis $H_1$ reflect the presence of a single point source located at $x=0$ while $H_2$ means that two half-brightness point sources are equally spaced about the origin with a separation of $\theta$. Based on the framework presented in Section~\ref{sec:O1}.\ref{sec:class}, the problem can be cast as discriminating between ${\hat \rho}_1^{\otimes N}$ and ${\hat \rho}_2^{\otimes N}$, where ${\hat \rho}_m$ is the density operator for each collected optical mode of the incoherent thermal field that contains one photon and $N$ is the total photon number collected during the optical recording time~\cite{lu2018quantum}. Using~\eqref{eq:singlephotonstate}, we have:
\begin{eqnarray}
{\hat\rho}_1 &=& |\psi_0\rangle \langle \psi_0|\label{rho1}\\
{\hat\rho}_2 &=&\frac12 
\bk{\ket{\psi_{\theta/2}}\bra{\psi_{\theta/2}}+\ket{\psi_{-\theta/2}}\bra{\psi_{-\theta/2}}}.\label{rho2}
\end{eqnarray}

The QCE in this case becomes
\begin{equation}
    \xi_{\rm Q}=-\log\left[\frac{1}{2}\left(\abs{\langle\psi_0\vert\psi_{\theta/2}\rangle}^2+\abs{\langle\psi_0\vert\psi_{-\theta/2}\rangle}^2\right)\right],
    \label{eq:xi_Q_1vs2}
\end{equation}
where we used~\eqref{eq:QCB_pure} for the special case when ${\hat \rho}_1$ is pure. The Chernoff exponent of SPADE from~\eqref{eq:CE} exactly matches its quantum counterpart (\ref{eq:xi_Q_1vs2}) for all values of $\theta$, confirming that it is an optimal measurement for 1-vs-2 point source discrimination just as it is for two-point-source separation estimation~\cite{lu2018quantum}. A Taylor expansion in $\theta$ yields the lowest order behavior of the QCE and SPADE CE, which is valid for any real-valued PSF~\cite{lu2018quantum}:
\begin{equation}
    \xi_{\rm Q}=\xi_{\rm SPADE}=\frac{(\theta\Delta k)^2}{4} + O\left((\theta\Delta k)^3\right).
    \label{eq:xi_Q}
\end{equation}
 On the other hand, Lu {\em et al.} calculated the classical CE of direct imaging with a Gaussian PSF to be~\cite{lu2018quantum}
\begin{equation}
    \xi_{\rm direct}=\frac{(\theta\Delta k)^4}{16} + O\left((\theta\Delta k)^6\right).
    \label{eq:xi_direct}
\end{equation}
Comparing~\eqref{eq:xi_Q} and~\eqref{eq:xi_direct} reveals a quadratic gap between the $\theta\Delta k$ scaling of the CEs for SPADE and direct imaging. This scaling gap means that the smaller the scene below the diffraction limit, the larger the extent to which SPADE can achieve a desired error probability with fewer photons (and hence less integration time) than conventional direct imaging~\cite{lu2018quantum}.

It is illuminating to quantitatively connect this result with the  point source separation estimation problem considered in Sections~\ref{sec:O1}. To do so, we consider a modified FI ${\tilde I}(\theta) \equiv \theta^2 I(\theta)$ that provides a lower bound on the variance of the \emph{relative} estimation error $\left(\check{\theta}-\theta\right)/\theta$~\cite{grace2022quantum}. This modified FI for SPADE  matches the modified QFI, evaluating as
\begin{equation}
    \tilde{I}_{\rm Q}(\theta)=\tilde{I}_{\rm SPADE}(\theta)=(\theta\Delta k)^2,
    \label{eq:QFI_rel}
\end{equation}
while the modified FI for direct imaging becomes
\begin{equation}
    \tilde{I}_{\rm Direct}(\theta)=2(\theta\Delta k)^4 + O\left((\theta\Delta k)^6\right)
    \label{eq:FI_direct_rel}
\end{equation}
We observe remarkable similarities between these results and those of Eqs.~(\ref{eq:xi_Q})-(\ref{eq:xi_direct}), namely the facts that SPADE saturates the quantum limit and that direct imaging exhibits a quadratic scaling deficiency with respect to the optimal measurement. These observations highlight the connection between estimation and hypothesis testing problems in our setting.


\subsection{How quantum is it?}
Before proceeding to describing modern research in this field, we address the following frequently asked question. Above, we derived the ultimate limits to sensing according to the quantum theory of light. However, the optical sources we consider are thermal, i.e.~``classical'' \cite{glauber1963quantum}. Their Glauber-Sudarshan $P$ function is positive, and hence their photodetection statistics can be modeled by semiclassical treatment of the photoelectric effect, in which quantum treatment is extended to matter only, and the light field is treated classically. Why do we apply quantum theory to phenomena that can be understood classically?

A simple answer is that measurements of optical states are not limited to detection via photoelectric effect. 
The quantum treatment allows us to find the ultimate limits for all possible measurements --- including those that cannot be explained semiclasically. Once these general limits are known, we can look for measurements that saturate them. Although in certain cases, as we found, such  measurements constitute linear optical processing followed by photodetection, initial quantum treatment is needed for making this determination. \blue{In other situations, particularly for multiparametric estimation discussed below, quantum-optimal measurements may require collective (entangling) measurements, in which case a fully quantum treatment of the light is indispensable.}


\section{Recent theoretical developments} \label{sec:O2}
The discoveries of Tsang \emph{et al.}\ have kickstarted a renewed interest in the quantum perspective of incoherent imaging. Their simplistic assumption of two equally bright sources, however, limits the application of the theory, and much theoretical work has been done since to generalize their results. Significant research has been devoted on the use of other types of sources, such as laser and nonclassical sources, for superresolution
imaging \cite{Kolobov00,genovese16,taylor16,FabreTrepsRMP,defienne2024}, although those studies are beyond the scope of this review. 

In this section, we will discuss some recent development in the theory of quantum-limited resolution, e.g., in multiple-hypothesis tests, accounting for partial coherence, effect of noise and imperfection in mode sorting, and various multi-parametric settings such as moment estimation, multi-source localization, estimating brightnesses and locations of point emitters, counting problems, imaging extended objects and accounting for nuisance parameters.

\subsection{Biased estimators}\label{sec:biased}
Despite the widespread acceptance of the Cram\'er-Rao bounds, one concern about them is that they rely on the assumption of unbiased estimators. This means that, if the estimator expectation value does not match the true parameter value, it may be able to violate the bounds. The behavior of biased estimators for two-point-source separation estimation was studied numerically in Appendix~E of Ref.~\cite{Tsang16} and experimentally in Ref.~\cite{Tham17}. A bias-corrected CRB is given by \cite{LehmannCasella}
\begin{align}
\MSE(\theta) &\ge \frac{[1+\dv*{b(\theta)}{\theta}]^2}{N I(\theta)}
+ b(\theta)^2,
\label{eq:unbiasedMSE}
\end{align}
where a scalar parameter $\theta$ is assumed and $b(\theta)\equiv \expect(\check\theta) - \theta$ is the bias of the estimator.  For the original CRB [$\MSE(\theta)\ge I(\theta)^{-1}/N$  as per Eq.~(\ref{CRB})] to be violated, $\dv*{b(\theta)}{\theta}$ needs to be negative. However, if this derivative stays significantly negative for a large range of $\theta$, the $b^2$ term in Eq.~(\ref{eq:unbiasedMSE}) would become large, causing the bias-corrected CRB
to go above the original CRB. \teal{As a result, biased estimators can offer improved performance only within a limited range of parameter values.}

\teal{For example, consider SPADE with the fundamental mode matching the PSF.} When $\theta = 0$, higher-order modes receive no photons --- so any estimator that returns $\check\theta = 0$ when no detection events are registered in higher-order modes is  perfectly correct and has zero MSE. For a nonzero but small $\theta$, the MSE will grow from zero \teal{but can remain below the lower limit for unbiased estimators given by QCRB (\ref{QCRB_sep}).}

\teal{This is the case for the maximum-likelihood estimator. For small $\theta$, the probability $N p(1|\theta)\approx N(\theta \Delta k)^2/4$ [as per \eqref{p1}]. For $(\theta \Delta k)^2/4\lesssim 1$, the most likely measurement outcome  is to observe zero photons in the first (and higher) excited modes, leading to the estimator returning a biased value $\check\theta=0$  --- so $b(\theta)=-\theta$ and $\dv*{b(\theta)}{\theta} \to -1$. 
As $\theta$ increases from zero, so does the MSE, eventually exceeding the QCRB. }

\begin{figure}[t]
	\centering
	\includegraphics[width=0.9\columnwidth]{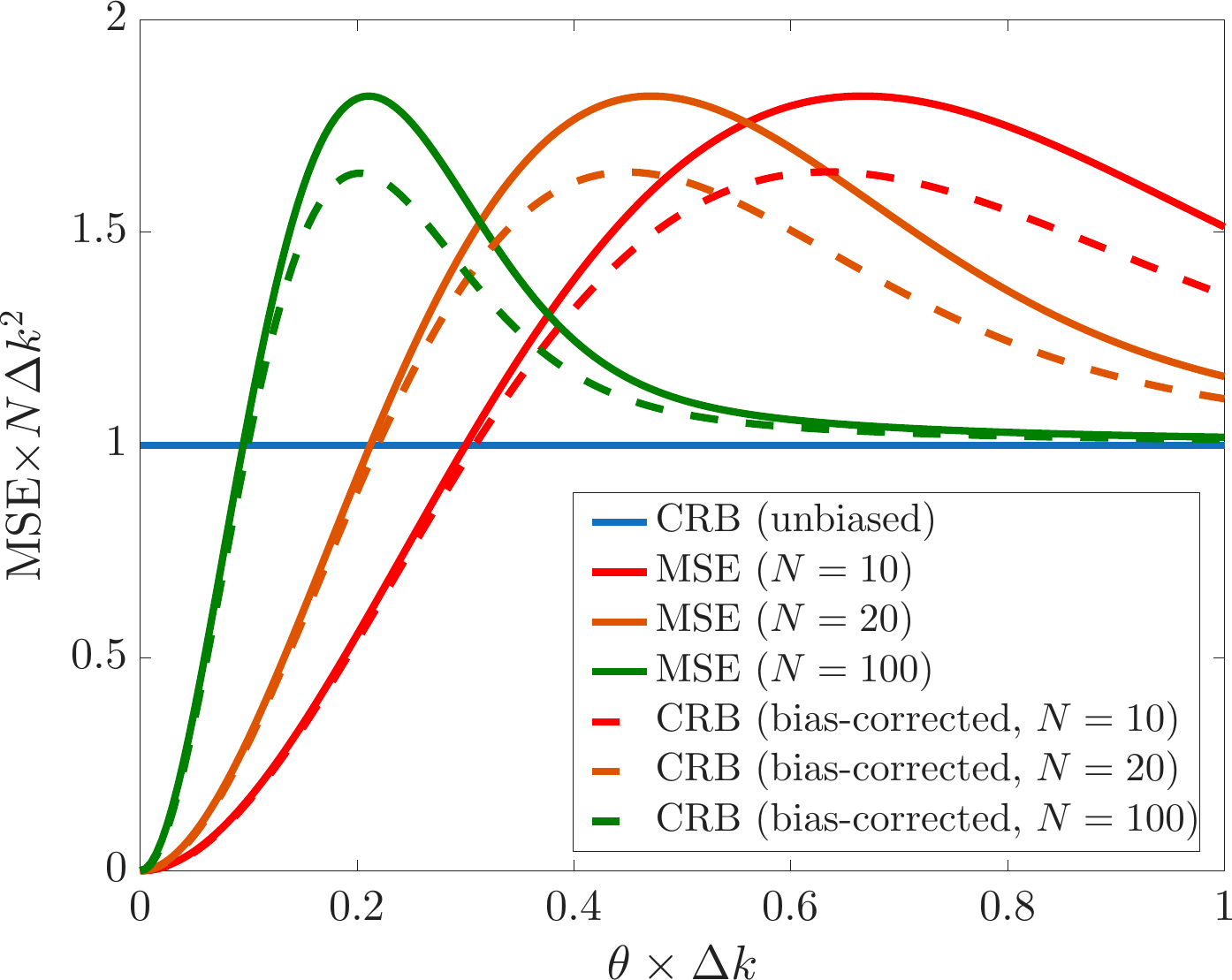}
	\caption{Numerically
    computed normalized errors $N\,\MSE(\theta) = N
    \sum_{n=0}^\infty [(2/\Delta k)\sqrt{n/N}-\theta]^2 e^{-N \Delta k^2\theta^2/4}(N\Delta k^2\theta^2/4)^n/n!$ of SPADE and maximum likelihood estimation for two point sources with a Gaussian PSF for different 
            detected photon numbers $N$,
            following Appendix~E in
            Ref.~\cite{Tsang16}. The dashed lines show the
            bias-corrected CRBs computed
            from~\eqref{eq:unbiasedMSE}. These can go below the CRB
            from~\eqref{CRB} because they account for the
            estimator-dependent bias present in the maximum-likelihood estimator with
            finite photon counts.  } 
    \label{fig:unbiasedMSE}
\end{figure}

\teal{This behavior is plotted in Fig.~\ref{fig:unbiasedMSE}, showing the normalized errors of SPADE with the maximum-likelihood estimator as well as the bias-corrected CRBs, demonstrating that the latter are indeed valid lower bounds on the respective MSEs at all $\theta$~\cite{grace2022quantum}.}  
We see that the region of $\theta$ in which $N\,\MSE(\theta) < N\,\CRB(\theta)$ grows smaller as $N$ increases, until $N\,\MSE(\theta) \to N\,\CRB(\theta)$ for all $\theta$ as the maximum-likelihood estimation becomes unbiased for $N\to\infty$ \cite{LehmannCasella}.

A more conclusive approach to remove the unbiased-estimator assumption is to assume that the parameter is sampled from some prior probability density $q(\theta)$. Then one can compute \emph{Bayesian lower bounds} on the average error
\begin{align}
\expect_q(\MSE) \equiv \int \MSE(\theta) q(\theta) d^n\theta,
\end{align}
which are proven to hold for any biased or unbiased estimator \cite{VanTrees2007bayesian}.  For a scalar parameter $\theta$, a Bayesian version of the CRB --- first proposed by Sch\"utzenberger \cite{schutzenberger57} and later independently by Van Trees \cite{VanTrees2013,VanTrees2007bayesian} --- is given by 
\begin{align}
\expect_q(\MSE) &\ge \frac{1}{N\expect_q(I) + J},
&
J &\equiv \expect_q\BK{\Bk{\pdv{\ln q(\theta)}{\theta}}^2},
\label{vantrees}
\end{align}
where $\expect_q(I)$ is the average Fisher information and $J$ quantifies the prior information.   A quantum version follows naturally from $I \le I_{\rm Q}$.

The Bayesian bounds are also lower bounds  on the worst-case error $\sup_\theta \MSE(\theta)$, since $\sup_\theta \MSE(\theta) \ge \expect_q(\MSE)$ for any prior.   
Using the bound in \eqref{vantrees} and picking the prior judiciously, Ref.~\cite{tsang18a} shows that, assuming a Gaussian PSF, the worst-case error for two-point separation estimation with direct imaging has a lower bound that scales as $1/\sqrt{N}$, that is,
\begin{align}
\sup_\theta\MSE^{(\textrm{direct})}(\theta) &\ge \frac{C}{\sqrt{N}},
\end{align}
where $C$ is some positive constant, whereas a quantum lower bound valid for any measurement is
\begin{align}
\sup_\theta\MSE(\theta) &\ge \frac{1}{4 N \Delta k^2},
\end{align}
which scales as $1/N$. Moreover, SPADE together with maximum-likelihood estimation is proved to achieve a worst-case error upper-bounded by twice the quantum bound, that is,
\begin{align}
\sup_\theta\MSE^{(\textrm{SPADE})}(\theta)  &\le \frac{1}{2 N \Delta k^2},
\end{align}
while a better estimator can further reduce the error. Hence, the superiority of SPADE still holds if we remove the unbiased-estimator assumption, although now the error is defined differently and the superiority is in terms of the photon-number scaling.

\subsection{Partially coherent sources}\label{sec:PartCoh}
Significant research has been dedicated to extending the point source separation estimation problem (Sec.~\ref{sec:O1}.\ref{sec:psse}) to coherent and partially coherent sources, \blue{which pertains to interferometric scattering microscopy and other settings in which laser illumination is involved, such as LIDAR.} These studies produced controversial results 
\cite{tsang15,larson18,tsang_comment19,lee19,kurd22,larson19,hradil19,liang21,wadood21}. In this subsection, we compute the QFI for this problem and explain the nature of the debate. 

For fully coherent sources, the simplest model of the incoming light field is the coherent state. 
If the amplitudes of the two sources are $\alpha_1$
and $\alpha_2$, the coherent state 
$\ket{\alpha(x|\theta)}$ in the image plane is characterized by the amplitude distribution
\begin{align}\label{alphaxtheta}
	\alpha(x|\theta) &= 
	\alpha_1 \psi(x-\theta/2) + \alpha_2 \psi(x+\theta/2).
\end{align}
To evaluate QFI, we compute the fidelity
\begin{align}
	\nonumber F &= \abs{\braket{\alpha(x|\theta)}{\alpha(x|\theta+\Delta\theta)}}^2
	\\ \nonumber 
	&= \exp\Bk{-\intall \abs{\alpha(x|\theta)-\alpha(x|\theta+\Delta\theta)}^2 
		\dd{x}}
	\\
	&\approx 1 -\frac{\Delta k^2\Delta\theta^2 }{4}
	\BK{|\alpha_1|^2+|\alpha_2|^2 -2\Re\Bk{\alpha_1\alpha_2^*
			\kappa(\theta)}} ,
\end{align}
where 
\begin{equation}
	\kappa(\theta) \equiv
	\frac{1}{\Delta k^2}
	\intall \pdv{\psi(x-\theta/2)}{x} \pdv{\psi^*(x+\theta/2)}{x}  \dd{x}
\end{equation}
is a function within the range $[-1,+1]$, taking the value of $1$ for $\theta=0$ and approaching zero at infinity. Accordingly, the QFI becomes
\begin{align}
	I_{\rm Q}(\theta) &= \Delta k^2
	\BK{|\alpha_1|^2+|\alpha_2|^2 -2\Re\Bk{\alpha_1\alpha_2^*
			\kappa(\theta)}}.
	\label{QFI_coh}
\end{align}

For partially coherent sources, we can treat the quantum 
state as a mixture of coherent states and $\alpha_1,\alpha_2$ as zero-mean random variables, with
their covariance determined by the mutual coherence \cite{Helstrom1976}.
The exact QFI then requires numerics to compute  \cite{kurd22,tsang_comment19,tsang21,wadood21}. However, owing to the convexity of the QFI with respect to the quantum state \cite{fujiwara01}, the right-hand side of Eq.~(\ref{QFI_coh}) gives an upper bound on the QFI if we replace $|\alpha_{1,2}|^2$ and $\alpha_1\alpha_2^*$ by their expectation values. For two equally bright sources ($\expect(|\alpha_1|^2) = \expect(|\alpha_2|^2)$), this result takes the simpler form 
\begin{align}
	I_{\rm Q}(\theta) &\le N \Delta k^2  \BK{1 -\Re\Bk{\gamma\kappa(\theta)}},
	\label{QFI_convex}
\end{align}
where $N/2 \equiv \expect(|\alpha_1|^2) = \expect(|\alpha_2|^2)$
and
\begin{align}
	\gamma &\equiv \frac{\expect(\alpha_1\alpha_2^*)}
	{\sqrt{\expect(|\alpha_1|^2)\expect(|\alpha_2|^2)}} = 
	\frac{2}{N} \expect(\alpha_1\alpha_2^*)
\end{align}	
is the complex degree of coherence satisfying $0\le|\gamma| \le 1$.

		\begin{figure}[t]
	\centering
	\includegraphics[width=\columnwidth]{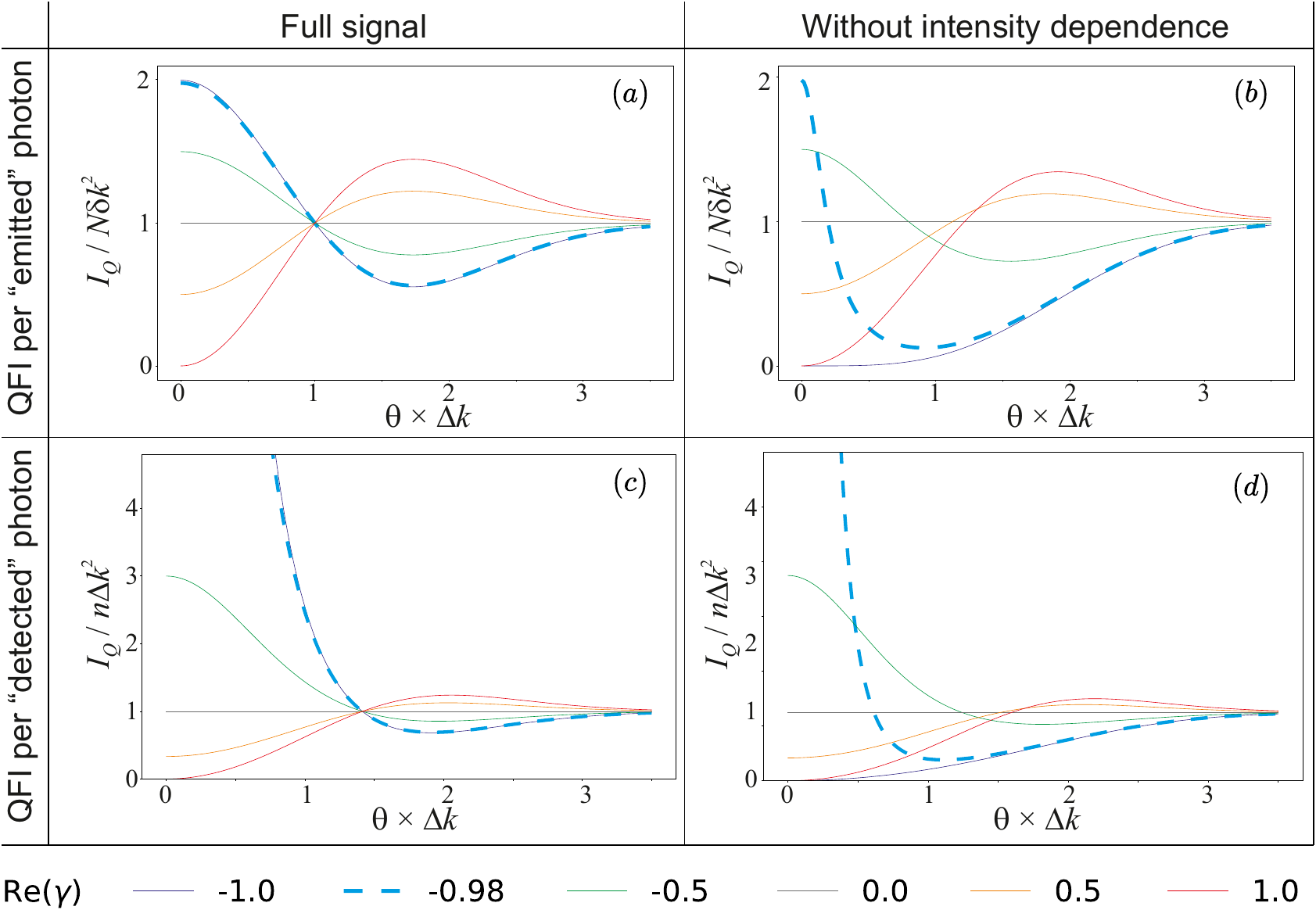}
	\caption{Per-photon QFI for Gaussian 
		Gaussian point-spread function \eqref{GaussPSF} for different values of $\textrm{Re}(\gamma)$. Top \emph{vs} bottom row represent to the QFI for emitted \emph{vs} detected photon. The right column calculates the QFI corresponding to the spatial shape of each photon, neglecting the additional information contained in the dependence of the number of defected photons on the source separation. Part (a), first  computed in Ref.~\cite{tsang_comment19} represents \eqref{QFI_convex}, whereas part (d) was obtained in Refs.~\cite{larson19,hradil19}. QFI for $\gamma=-1$ in the second column is singular, which is illustrated by the stark difference between $\textrm{Re}(\gamma)=-1$ and $\textrm{Re}(\gamma)=-0.98$. Reproduced from Ref.~\cite{kurd22}, modified to adjust to this paper's notation.	}		\label{kurdfig}
\end{figure}

The upper bound (\ref{QFI_convex}) is plotted for a one-dimensional, Gaussian PSF in Fig.~\ref{kurdfig}(a). We see that Rayleigh's curse is present (i.e.~$I_{\rm Q}(0) = 0$) only for coherent in-phase sources ($\gamma=1$), whereas the QFI for coherent out-of-phase sources ($\gamma=-1$) is twice that for the incoherent case ($\gamma=0$).  
This behavior can be intuitively understood by considering the FI (\ref{derivmode}) due to the derivative mode (Fig.~\ref{spade_explain1}). The intensity of this mode in the low-intensity limit is 
\begin{align}\label{p1coh}
	p(1|\theta) &\propto
	\expect
	\BK{\abs{\intall \psi_1^*(x)\Bk{\alpha_1 \psi(x-\theta/2)+\alpha_2\psi(x+\theta/2)} \dd{x}}^2}.
\end{align}
In contrast to the incoherent case (\ref{p1}),  the two contributions  now interfere. In the sub-Rayleigh regime $\theta\ll 1$, the amplitudes due to the two sources are $\propto -\alpha_1\theta/2$ and $\propto \alpha_2\theta/2$, respectively, so $p(1|\theta) \propto \expect(|-\alpha_1\theta + \alpha_2\theta|^2)\propto 1 - \Re\gamma$. In particular, when the sources are coherent and in-phase ($\alpha_1=\alpha_2$), the transverse mode of the field is symmetric and does not contribute to the derivative mode, giving rise to the Rayleigh curse. On the other hand, for coherent out-of-phase sources, the Fisher information is twice that of the incoherent case. 

A remarkable additional result is obtained from a more accurate calculation of the integral (\ref{p1coh}):
\begin{align}
	p(1|\theta) \approx \frac{N\Delta k^2 \theta^2}{4} \bk{1 - \Re\gamma},
	\textrm{ so }
	I(\theta) \approx N\Delta k^2\bk{1 - \Re\gamma}.
\end{align}
We see that the upper bound (\ref{QFI_convex}) is in fact exact in the sub-Rayleigh regime and SPADE can attain the quantum limit for any
$\gamma$. A more exact analysis of SPADE reveals that the
attainment is perfect for any $\theta$ as well as any $\gamma$
\cite{kurd22}.

As mentioned, different studies of partially coherent sources
\cite{larson18,larson19,hradil19,liang21} have produced different
results. Kurdzia{\l}ek \cite{kurd22} analyzed these differences in detail and showed that they arise when one attempts to estimate the QFI \emph{per photon}. In contrast to the incoherent case, the number of photons in the image plane, obtained from \eqref{alphaxtheta} by integrating 
\begin{equation}
	n =\int\limits_{-\infty}^\infty|\alpha(x|\theta)|^2\dd x
\end{equation}
depends on the source separation $\theta$. Accordingly, the QFI (\ref{QFI_convex}) can be prorated either by the number $N$ of ``emitted'' photons, giving rise to Fig.~\ref{kurdfig}(a) or by the number $n$ of ``received'' photons, leading to Fig.~\ref{kurdfig}(c). In the latter case, the setting $\gamma=-1$ (coherent sources with opposite phases) leads to $n\to0$ for $\theta\to 0$ due to destructive interference. Accordingly, the QFI per received photon tends to infinity.

Importantly, the dependence of $n$ on $\theta$ inherently carries information about this parameter. If one ignores this dependence, choosing to only analyze the spatial wavefunction of each arriving photon, one will obtain the set of curves shown in Fig.~\ref{kurdfig}(b,d). \blue{This situation would also arise if the observer is unaware of the sources's amplitudes.} Because \blue{in this case only a part of the} information carried by the optical field \blue{is utilized}, the QFI in this case does not exceed its counterpart for the complete signal [Fig.~\ref{kurdfig}(a,c), respectively]. An exception is  $\gamma=-1$, which leads to a singular expression for the QFI
and therefore cannot be reliably analyzed \cite{hradil19}. \blue{Aside from this ``pathological'' case, Rayleigh's curse remains limited to the case of in-phase sources ($\gamma=1$). Additionally, Rayleigh's  curse is present for all relative phases if the degree of coherence is unknown \emph{a priori} \cite{larson19}. This can be of practical relevance if the sources interact with each other so their mutual coherence is affected by their relative distance.}

In our view, these disagreements can be resolved by recognizing that the QFI is an intrinsic property of the quantum state and not necessarily proportional to the number of photons contained in that state. Hence, the very notion of per-photon QFI \teal{ for states containing coherent superpositions of photon numbers} may be questionable \cite{yang2017}.

\subsection{Hypothesis testing for generalized objects}
\label{sec:HypothesisTesting}
After Refs.~\cite{Krovi2016,Lu2016,lu2018quantum} showed that SPADE can improve discrimination between symmetric point objects, the next step was to investigate hypothesis \magenta{tests} of real-world interest. 
Huang and Lupo~\cite{huang2021quantum} considered \magenta{exoplanet} detection as an asymmetric test between the hypothesis that no \magenta{exoplanet} is present, so \magenta{that} the state is given by the single-photon PSF (\ref{rho1}), and the star-exoplanet hypothesis, corresponding to the state
\begin{equation}
    \hat{\rho}_2=(1-b)\ket{\psi_0}\bra{\psi_0}+b\ket{\psi_\theta}\bra{\psi_\theta}, \label{exoplanetHypothesis}
\end{equation}
where $b$ is the relative brightness of the exoplanet compared to the total star-exoplanet system. The authors bounded the probability of a false negative result (missing an exoplanet) under a constrained false positive rate using the quantum Stein lemma~\cite{Ogawa2000}. This led them to analyze the (quantum) relative entropy, which is an extension of the (quantum) Chernoff exponent to asymmetric hypothesis tests, when errors for each of the two hypotheses are assigned unequal costs. For a Gaussian PSF, they computed the quantum relative entropy to be $D(\hat{\rho}_1||\hat{\rho}_2)=\left(1-\exp(-\theta^2\Delta k^2/16)\right)b+O(b^2)$ and showed that SPADE achieves this bound for the practically relevant case of $b\ll1$. The relative entropy produced by direct imaging was $D(p_1||p_2)=(1/2)\left(\exp(\theta^2\Delta k^2/4)-1\right)b^2+O(b^3)$, a linear factor of $b$ below the quantum limit. Deshler \emph{et al.} later found that the same scaling gap with respect to exoplanet brightness appears in the (Q)CEs for symmetric hypothesis tests in the $b\to 0$ limit, and also that modern coronagraphs achieve the optimal scaling $D(p_1||p_2)\sim c b$  with a sub-optimal prefactor $c$~\cite{Deshler2025}.

\begin{figure}[t]
	\centering
	\includegraphics[width=0.8\columnwidth]{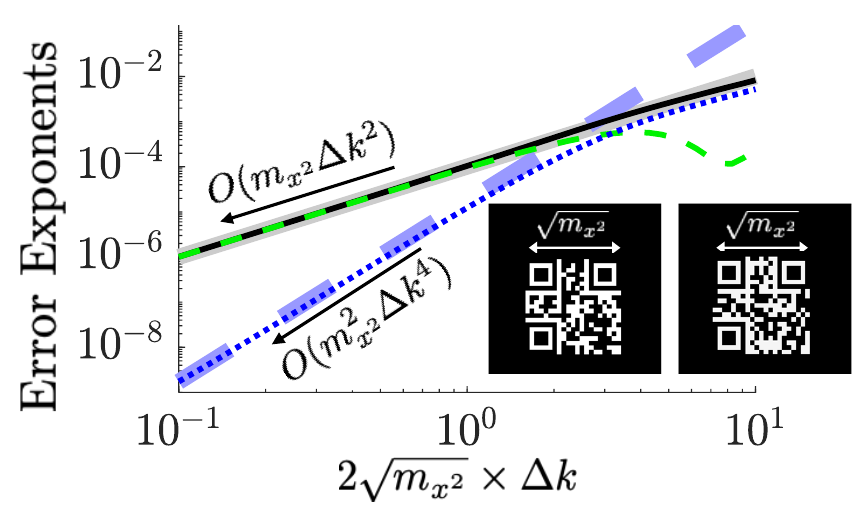}
	\caption{Chernoff bounds on error exponents for discrimination between two QR codes (inset) with a 2D Gaussian aperture. The CE of TriSPADE (green) converges to the QCE (black) in the sub-Rayleigh limit $\sqrt{m_{x^2}}\Delta k \ll 1$. The CE of direct imaging (narrow blue) decreases quadratically faster with respect to $\sqrt{m_{x^2}}\Delta k$ than the quantum limit. \magenta{Analytical lowest-order approximations for the QCE (thick gray) and direct imaging CE (thick blue) closely match numerical results.} \blue{Adapted from~\cite{grace2022identifying}. Copyright 2022 American Physical Society.}
    }
    \label{fig:hypothesistest_results}
\end{figure}

\magenta{Generalizing the class of objects being identified,} Grace and Guha~\cite{grace2022identifying} computed (quantum) Chernoff bounds for \magenta{libraries} of \blue{$M$} objects with 2D intensity profiles $I_j(\boldsymbol{r})$, $1 \le j \le M$, finding that the bounds depend primarily on the object second moments $m_{j,a^2}\equiv\iint_{\mathbb{R}^2}a^2I_j(x,y)dxdy$ for $a\in\{x,y\}$. 
\magenta{When the intensity distributions of two objects have coinciding angular axes of principal variance~\cite{triggiani2026achieving}, Ref.~\cite{grace2022identifying} calculated (Q)CEs} to lowest order in the resolvability measure $\sqrt{m_{j,a^2}}\Delta k$,
where $\Delta k\equiv\left[\iint_{\mathbb{R}^2}\abs{\partial\psi(a)/\partial a}^2 dxdy\right]^{1/2}$ is the extension of \eqref{Deltak} to two dimensions. 
For a 2D Gaussian PSF, the authors \magenta{proved a quadratic scaling gap between $\xi_{\rm Q}=O(m_{j,a^2}\Delta k^2)$ and $\xi_{\rm direct}=O(m_{j,a^2}^2\Delta k^4)$, generalizing the result from the one-vs-two point source problem (see Eqs.~(\ref{eq:xi_Q}) and (\ref{eq:xi_direct})). This implies} a dramatic improvement in hypothesis testing accuracy \magenta{over direct imaging} for sub-Rayleigh scenes.  
\magenta{To manifest this advantage, the authors proposed a simplified SPADE receiver,} dubbed TriSPADE, \magenta{that detects three HG modes:} $(0,0)$, $(0,1)$ and $(1,0)$. \magenta{They showed that numerically calculated CEs for TriSPADE converge to the QCE and quadratically outperform direct imaging} for a variety of objects, including quick response (QR) codes \magenta{(Fig.~\ref{fig:hypothesistest_results})}.

\magenta{Triggiani and Lupo identified two sufficient conditions for TriSPADE to asyptotically achieve the QCE in the sub-Rayleigh limit: both objects have the same angular axis of principal variance, and the TriSPADE receiver is rotated to match that axis~\cite{triggiani2026achieving}. \blue{The first condition is satisfied for objects} 
with no expected reference angle (e.g., freely diffusing biostructures or orbital satellites); since there is only one object in the scene, there is no explicit relative angle between candidates, 
and one can assign the same principal axis to both objects. 
\blue{Even when this condition is not satisfied, the sub-optimality of TriSPADE is insignificant in certain important scenarios.} For example, while the QR codes in Fig.~\ref{fig:hypothesistest_results} have different principal axes, the CE for TriSPADE (green) nearly matches the QCE (black). Satisfying Triggiani and Lupo's second condition without prior knowledge of the principal axis remains an open question. One possible approach is that of Ref.~\cite{grace_approaching_2020}, where a preliminary measurement(s) estimates the object location and angle in order to align and rotate a TriSPADE.}

\magenta{These results extend} to larger object libraries ($M > 2$)~\cite{grace2022identifying}. For equiprobable objects, the $M$-ary QCE $\xi_{{\rm Q},M} = \min_{i\ne j}\xi_{{\rm Q},(i,j)}$ gives the optimum error exponent in discriminating between $N$ copies of states ${\hat \rho}_i$, $1 \le i \le M$, where $\xi_{{\rm Q},(i,j)}$ is the pairwise QCE for ${\hat \rho}_i$ and ${\hat \rho}_j$~\cite{Li2016}. The import of this result is that the hardest-to-discriminate pair determines the overall error exponent.


\subsection{Multiparameter estimation}

When considering the simultaneous estimation of multiple parameters, the bound in (\ref{CRB}) is not guaranteed to be tight, meaning that a single  measurement yielding the optimal simultaneous estimation of all parameters may not exist in general, as discussed in Sec.~\ref{sec:O1}.\ref{sec:est} \cite{Ragy2016}. \teal{Furthermore, the uncertainty boundary on an individual parameter $\theta_i$ cannot be estimated as the inverse of the FI matrix element $1/(I)_{jj}$ corresponding to that parameter; rather, it is the matrix element $(I^{-1})_{jj}$ of the inverse of the entire FI matrix.} The inequality $(I^{-1})_{jj} \geq 1/I_{jj}$ \cite{Helstrom1976} implies that statistical correlations among the parameters  reduce the individual precision on estimating each parameter when the others are unknown. Below, we review some case studies based on applications of multiparameter estimation theory.

\subsubsection{Unknown or misaligned centroid}\label{sec:centroid}
\begin{figure}[t]
\centering
\includegraphics[width=0.9\linewidth]{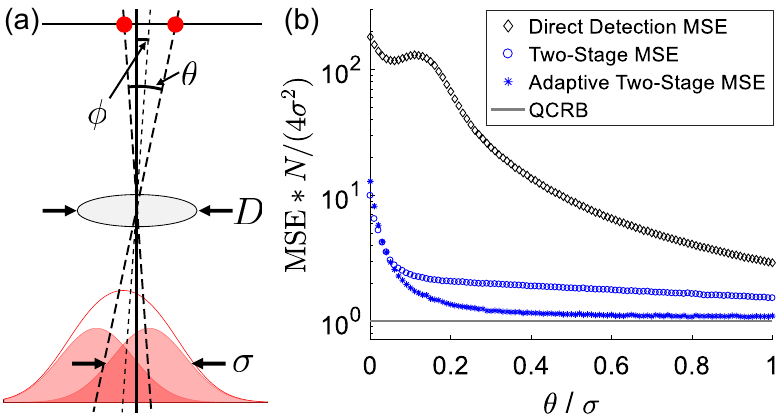}
\caption{(a) Schematic of measuring the two-point source separation $\theta$ with unknown centroid $\phi$ imaged with a Gaussian aperture. (b) Per-photon simulated MSE for the estimation of $\theta$ with $\sigma=1/2\Delta k$. Reproduced from Ref.~\cite{grace_approaching_2020}}.
\label{fig:centroid}
\end{figure}

A common confounding feature of multiparametric imaging scenarios is the presence of "nuisance parameters," which are defined as parameters of the inference model that are not explicitly required to be estimated but that can impede sensitivity if ignored. The earliest analysis of a nuisance parameter for SPADE-like imaging receivers considered 1D spatial misalignment of the receiver's optical axis with respect to the centroid of two identical point sources [Fig.~\ref{fig:centroid}(a)]. It was found that SPADE is highly sensitive to this misalignment~\cite{Tsang16,zhou19}, which can be conceptually understood from Fig.~\ref{spade_explain1}. Under spatial misalignment, the first-order terms in the Taylor expansion no longer \teal{cancel each other, so signal is present even for $\theta=0$.} 
As a result, the constant QFI for sub-diffraction $\theta$ observed under perfect alignment is not attainable when misalignment is present, even when that misalignment is known~\cite{Tsang16}; the attainable FI for SPADE-like measurements turns out to scale as $O(\theta^2)$ for sub-diffraction point source separation estimation~\cite{zhou19}. Ref.~\cite{de2021discrimination} also showed that a similar scaling suboptimality is observed for SPADE with respect to the QCE for one-vs-two point-source hypothesis testing. These results all hold in the case where the misalignment is static for the entire measurement duration, which is the most naturally occurring scenario. Interestingly, if the misalignment is a zero-mean stochastic variable that changes faster than the sampling rate of the photodetectors, then SPADE becomes a quantum optimal measurement once again ~\cite{grace2022quantum}.

Despite the detrimental effects of static misalignment, advantages over direct imaging are still possible with practical SPADE receivers. If the degree of misalignment is known \emph{a priori}, the SPADE mode set can be displaced to compensate for this misalignment. However, even if this correction is impossible due to technical limitations, SPADE can be used with no changes to achieve a constant factor advantage in estimation precision over direct imaging~\cite{Tsang16}. 

When the initial misalignment is a nuisance parameter that is not known \emph{a priori}, \teal{it can be estimated by direct imaging. This leads to the problem of measurement budget allocation: what fraction of measurements should be done in the direct imaging regime to estimate the centroid, with the remainder of the measurement to be done in the HG SPADE basis. For the deep sub-Rayleigh case Ref.~\cite{zhou19} proved that a predetermined equal time-allocation ratio between direct imaging and SPADE is optimal. Outside the deep sub-Rayleigh regime,} Ref.~\cite{grace_approaching_2020} proposed an adaptive algorithm in which the total available integration time is divided between a preliminary direct imaging measurement, which is used to estimate the centroid of the point source constellation, and a SPADE receiver that is adaptively aligned to the centroid estimate. The time allocation ratio between the two measurement stages was also dynamically optimized in an online fashion, where the algorithm periodically checks whether it is informationally favorable to acquire more direct imaging data or switch to the SPADE measurement.  This adaptive scheme was shown in simulation to achieve a 100x improvement over direct imaging in mean squared error for estimating both sub-diffraction point source separations as well as the size of sub-diffraction extended objects [Fig.~\ref{fig:centroid}(b)].

Regarding the precision of pre-estimating the centroid of a scene, \teal{Tsang {\em et al.} showed that image-plane direct imaging is QFI-suboptimal for the two-source constellation~\cite{tsang2016quantum}, but approaches the QFI {\em both} in the far-sub-Rayleigh and large-separation limits. This conclusion was extended by Sajjad et al.~\cite{Sajjad2021} to multiple point source constellations. They found that a two-stage adaptive SPADE measurement can in general attain the QFI of centroid estimation for any constellation --- suggesting thereby that the use of direct imaging as a precursor to an HG SPADE for the unknown-centroid two-source constellation may not be the optimal approach.}
However, since direct imaging is always asymptotically optimal for centroid estimation in the deep sub-Rayleigh regime, it serves as a reasonably good pre-calibration measurement for multi-stage adaptive receivers~\cite{grace_approaching_2020,lee2022quantum}.



Finally, substantial benefit can be gained on optimally estimating both the centroid and separation by making use of a qubit model for subRaleigh imaging that maps these two parameters to the rotation and contraction of a vector on the Bloch sphere, respectively~\cite{Chrostowski2017}. By working with this qubit model, several novel receiver strategies have been proposed to improve the Fisher information advantage over direct imaging without requiring a prior estimate of the centroid. These receiver designs include post-processing the optical outputs of a misaligned SPADE~\cite{de2021discrimination}, measuring the spatially resolved two-photon interference pattern of the incoming light~\cite{parniak2018beating}, and higher-order multi-photon collective measurements~\cite{de2025superresolving}.




\subsubsection{Power-imbalanced 2 sources}\label{sec:PowerImbalanced}
\teal{Even more interesting physics arises if, in addition to the centroid and separation of the two sources, a third unknown parameter is introduced: variable relative brightness of the sources. The QFI matrix for this setting was found by \v{R}eha\v{c}ek {\em et al.}~\cite{Rehacek17}. They found that the SLDs for the brightness and separation weakly commute, meaning that a (in general, a collective) measurement achieving QCRB for all of them exists in principle. Such a measurement was found theoretically by de Almeida {\it et al.}~\cite{de2025superresolving} --- a collective measurement on $N$ temporal modes requiring coupling the quantum state of the light into logic-capable atomic quantum memories. Practical realization of such a measurement is however challenging at the current level of quantum technology. }

A more practical solution was proposed by Ref.~\cite{Deshler2025}, inspired by super-resolution magnetic-field imaging using nitrogen-vacancy (NV) centers in diamond --- where the brightness of an emitter encodes the local magnetic field strength. They put forward an adaptive SPADE protocol containing three sets of measurements: direct imaging (optimized for centroid estimation), followed by the HG SPADE (optimum for separation estimation), and then followed by what they define as the Yuen-Kennedy-Lax (YKL) SPADE which borrows ideas from quantum-state discrimination theory (optimized for relative brightness estimation). This protocol achieves superior performance for relative brightness estimation compared to conventional direct imaging. 

Exoplanet detection is a particularly important use case of the power-imbalanced setting when the imbalance is extreme ($b_1 \approx 1$). In this scenario, both for the estimation problem (of estimating the value of the separation $s$) as well for the hypothesis-test problem (that of guessing whether or not a dim source is present next to the bright source), a receiver that performs image-plane direct imaging after removing the content of the image-plane field in the centered PSF mode, was found to be quantum (QFI and QCE) optimal~\cite{Deshler2024}.

\subsubsection{Three-dimensional localization}\label{sec:3Dloc}

Several studies have generalized the planar analysis to consider the simultaneous estimation of both lateral and axial components of the separation between point sources \cite{Ang17,Napoli19,YuPrasad2019,PrasadYu2019,Fiderer2021,Wang21,backlund2018fundamental,Frank2026,Muller2026}. 
For two equally bright incoherent point sources in the paraxial approximation, regardless of the specifics of the PSF, if one assumes that the centroid coordinates are known, then the QFI matrix for estimating all three-dimensional components of the separation vector $\boldsymbol{\theta}=(\theta_x,\theta_y,\theta_z)$ is diagonal and independent of such separations \cite{Ang17,Napoli19,YuPrasad2019,PrasadYu2019,Fiderer2021,Wang21}. This follows because the QFI matrix  can be expressed simply in terms of the correlation of the wavefront phase gradients in the imaging aperture, and the wavefront phase is linear with respect to the separation vector \cite{YuPrasad2019,PrasadYu2019,Fiderer2021} \purple{[GA to improve sentence]}. Specifically, for the Gaussian aperture we have \cite{Ang17,Napoli19,Fiderer2021}
\begin{align}\label{eq:QFI3D}
\left[I_{\rm Q}(\theta_x,\theta_y,\theta_z)\right]^{-1} = \textrm{diag}\left(\frac{1}{\Delta k^2}, \frac{1}{\Delta k^2}, \frac{k^2}{\Delta k^4}\right)\,,
\end{align}
which directly generalizes the one-dimensional QCRB (\ref{QCRB_sep}).



These bounds can in principle be achieved by physically realizable measurements. The $x$ and $y$ components of the separation vector are obtained by straightforward generalization of SPADE to two dimensions \cite{Ang17}. The axial component $\theta_z$ is encoded differently: it appears as a defocus-induced curvature of the optical wavefront. Yu and Prasad~\cite{YuPrasad2019} showed that optimal measurements correspond to projections onto modes whose transverse profiles are proportional to the derivative of the PSF with respect to defocus. For a circular pupil these modes are closely related to low-order Zernike polynomials, 
while for Gaussian pupils, the defocus derivative is expanded in radial Laguerre–Gaussian modes~\cite{zhou2019quantum,Muller2026}.

A complementary approach uses interferometric dual-objective microscopes. Backlund \textit{et al.}~\cite{backlund2018fundamental} analyzed systems in which fluorescence collected by two opposing objectives is combined coherently. The interference signal depends on the optical path difference and therefore encodes the axial position. 
Under shot-noise-limited detection, the achievable localization precision is consistent with the fundamental quantum limits.

More general settings involving two and three sources were considered theoretically in \cite{Fiderer2021,Frank2026}, including the simultaneous seven-parameter estimation of the centroid coordinate vector, the separation vector, and the relative intensity of two incoherent sources with uneven brightness. Quantum precision bounds for these settings were derived exactly thanks to a general analytical method to calculate the QFI matrix \cite{Napoli19,Fiderer2021}.

\subsubsection{Multi-point-source localization} \label{sec:MultiLoc}
Multiple studies considered the use of modal imaging for two-dimensional localization of a scene comprised of an unknown number of multiple point sources~\cite{Bisketzi_2019,bao21,lee2022quantum,matlin22,choi24}, possibly of different brightnesses, but with no clutter, i.e., with a guarantee that no other object apart from the point sources of interest are present in the field of view. For example, Lee {\em et al.} \cite{lee2022quantum} employed an adaptive mode sorter whose modal bases were evaluated using a Bayesian update scheme. 
\teal{The localization results from Fig.~\ref{fig:multisource_results} show accurate performance of this method, in contrast to ideal direct imaging, which localizes three point sources as a single emitter at the centroid of the true emitters.} \purple{[SG: Figure needs improvement.]} 

{\color{red}
	 \begin{figure}[t]
	 	\includegraphics[width=\columnwidth]{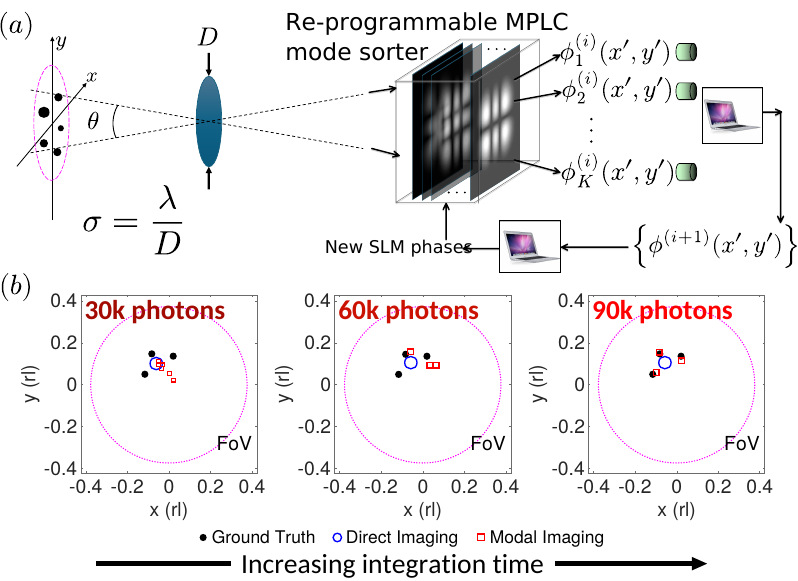}
	 \caption{(a) Dynamically programmable SPADE shown to image a collection of point emitters. (b) Simulation results: black dots denote ground-truth emitter locations inside a $0.8 \times 0.8$ Rayleigh-Unit field of view \blue{(denoted as rl and defined in Ref.~\cite{lee2022quantum} as the FWHM of the amplitude PSF: $\mathrm{rl}=4\sigma\ln 2$).} Blue circles denotes the estimate of the location (diameter denotes estimated brightness) generated by image-plane direct imaging, which fails to resolve the emitters even with $N=90,000$ collected photons; red squares denote emitter-location estimates generated by the Bayesian adaptive modal imaging algorithm.}
     \label{fig:multisource_results}
	 \end{figure}
}

\subsubsection{Moments estimation and towards full imaging}\label{sec:moments}
In addition to parameter estimation and hypothesis testing, substantial effort was directed toward full quantitative image reconstruction of sub-diffraction scenes with resolution surpassing that allowed by diffraction-limited imaging systems. Yang {\it et al.}~proposed extension of SPADE to this application \cite{Yang16}. They considered a coherent object with the field  distribution in the object plane given by $E(x,y)$. Assuming a Gaussian PSF (\ref{GaussPSF}), and the exponent in the HG mode basis matching that PSF, one can show that the amplitude of the HG$_{mn}$ component of the field in the image plane equals
\begin{align}\label{eq:HGM_J_n}
	J_{mn}
 = \frac{1}{\sqrt{m! n!}}\iint E(x,y)\left(x\Delta k\right)^m \left(y\Delta k\right)^n e^{-\Delta k^2(x^2+y^2)/2^2} \dd{x} \dd{y}.
\end{align}
When this distribution is sufficiently narrow, $J_{mn}$ approximates its $(m,n)$ th moment. The knowledge of all moments of a distribution with a finite support allows one, in principle, to reconstruct the distribution itself. The caveat is that, for higher mode orders, the mean photon number coupled to each mode decreases substantially, so a much longer integration time is required to estimate a higher-order moment of a subdiffraction object accurately. 

The method can be readily extended to incoherent imaging. For $I(x,y)$ being the intensity distribution of the incoherent object, the intensity (square absolute value) of the HG$_{mn}$ component of the image field is
\begin{align}\label{Pmn}
    P_{mn} &=  \frac{1}{m! n!} \iint I(x,y) \left(x\Delta k \right )^{2m} \left(y\Delta k \right )^{2n} e^{-\Delta k^2(x^2+y^2)} \ \dd{x} \dd{y},
\end{align}
i.e.~approximates the $(2m,2n)$th moment of the source field \cite{Yang16}. 

To obtain an intuition, let us specialize to $(m,n)=(1,0)$ and consider again \eqref{p10}. We see that the signal $P_{01}$ from a point source into the derivative mode HG$_{10}$ is proportional to the brightness of the source times the square of its $x$ coordinate. For multiple sources, the total energy in the derivative mode is the sum (integral) (\ref{Pmn}) of individual contributions --- that is, the $(2,0)$th moment of the intensity distribution. 



The fact that only the even moments can be measured means that only the even component of the source intensity distribution, given by $[I(x,y)+I(-x,y)+I(x,-y)+I(-x,-y)]/4$, can be reconstructed from these measurements. Yang {\it et al.} proposed to address this complication by displacing the object from the optical axis of the measurement apparatus, so the four terms of the above expression are separated in space . This approach is however suboptimal because it required a larger number of modes to cover an object of a large effective spatial extent and, furthermore, cannot be applied if the object is large or infinite in size, so its displacement will not separate the four terms.

An alternative solution was proposed by Tsang \cite{Tsang17}: to reconstruct the odd moments by measuring in an overcomplete basis, which includes not only HG modes, but also superposition of HG modes with neighboring indices. An alternative overcomplete basis, consisting of two sets of HG modes rotated by $45^\circ$ with respect to each other, was proposed by Buonaiuto and Lupo \cite{buonaiuto2024sub}. For an arbitrary centrosymmetric PSF, the TEM basis can be generalized to the so-called point-spread-function-adapted basis \cite{Rehacek17.1}, which can be used similarly for moment estimation \cite{Tsang18}. Furthermore, if the scene is comprised of fluctuating point sources (as in many fluorescence microscopy settings), it was recently shown that all moments of a 1D scene can be estimated from the temporal cumulants of the fluctuating fields measured from the first two spatial modes~\cite{kurdzialek2025}.

It can be shown that the performance of SPADE for moment estimation is far superior to that of direct imaging for moments of order $p + q \ge 2$ and objects with subdiffraction sizes \cite{Tsang17,Tsang18}. \teal{This is because, as shown above, each moment directly corresponds to a particular element of the HG measurement basis --- therefore optimizing the evaluation precision for the moments.}

The computation of classical and quantum bounds for moment estimation is challenging because the number of unknown parameters is infinite for an arbitrary distribution, even though the parameter of interest, a moment in this case, is a scalar function of this infinite parameter set.  In statistics, the estimation of a parameter of interest in the presence of an infinite number of unknown parameters is called semiparametric, and it is possible to define and compute the Cram\'er-Rao bound for semiparametric problems \cite{Bickel93,tsang19a}. By generalizing the classical approach, a semiparametric QCRB can be defined rigorously \cite{TsangAlbarelliDatta20} and evaluated for moment estimation. The theoretical and numerical results obtained this way \cite{tsang21a,tan23} agree with earlier explorative studies \cite{zhou19,Tsang19Imaging}, showing that SPADE is close to quantum-optimal for moment estimation. Ref.~\cite{tsang23} relates the moment parameters to generalized Fourier coefficients of the source distribution, showing that SPADE can also enhance the estimation of the Fourier     coefficients. 

\teal{In a related study,}  Dutton {\em et al.} considered the problem of estimating the length of a uniformly-bright line-shaped object with a known centroid, which they approached by a linear-constellation of $M$ equally-bright point emitters located in $[-\theta/2, \theta/2]$, and took the limit $M \to \infty$. Like the $M=2$ case studied by Tsang {\em et al.}~\cite{Tsang16}, the direct imaging FI crashes as $\theta$ dips below the Rayleigh limit and goes to $0$ as $\theta \to 0$. On the other hand, the QFI for estimating the length of the object $\theta$, assuming a Gaussian PSF, is exactly attained for all $\theta$ by the HG-SPADE. Unlike the $M=2$ case, the QFI per photon was seen to slowly decrease as $\theta$ increases, which is attributable to the fact that photons emanating from the middle of the line object are less informative about $\theta$ compared to the photons emitted from the end points, and hence as $\theta$ increases, a progressively larger fraction of the photons carry less QFI per photon. 

\blue{General 2D imaging --- reconstructing the object intensity $I(x,y)$ based on its diffraction-limited far field --- has for a long time been an uncharted territory in quantum sensing. This is because this intensity has infinitely many degrees of freedom, so the estimation problem has no natural finite parametrization. Accordingly, it is challenging to construct a quantum bound on image reconstruction precision and find a measurement that saturates this bound. Two recent works address this by different finite parametrizations, then evaluate the quantum limit and construct a measurement attaining it. Warke \emph{et al.}~\cite{warke26} describe the object by the amplitudes of its spatial-frequency components: since the aperture transmits nothing above the incoherent cutoff, truncating the Fourier expansion at theis point discards no observable information, therefore leaving a finite parameter set. 
Deshler \emph{et al.}~\cite{deshler26} instead approximate the scene by a grid of thermal point
emitters at known positions whose brightnesses are the unknowns. In both cases the SLDs of the
parameters weakly commute making QCRB theoretically achievable. However, under non-entangling (single-copy) measurements, the relevant limit is the tighter
Nagaoka--Hayashi bound~\cite{conlon2021efficient,tsang26,zhou26}, computed in both works by semidefinite
programming. Deshler
\emph{et al.} show a protocol for constructing adaptive  SPADE basis whose precision approaches that bound. 
Warke \emph{et al.} go further and specify a physical apparatus: a multiplane light converter (see Sec.~\ref{sec:O3}.\ref{sec:MPLC}), whose masks are trained by gradient descent to minimize the trace of the CRB matrix,  followed by an array of photon detectors. In this way, the optimal measurement basis emerges as a by-product rather than having to be found and then engineered. The trained network saturates the Nagaoka--Hayashi bound across the transmitted band in one and two dimensions, and
recovers sub-Rayleigh features that direct imaging with deconvolution does not.}

\section{Structured receivers for sub-diffraction optical sensing}
\label{sec:O3}

\begin{figure}[t]
	 \includegraphics[width=\columnwidth]{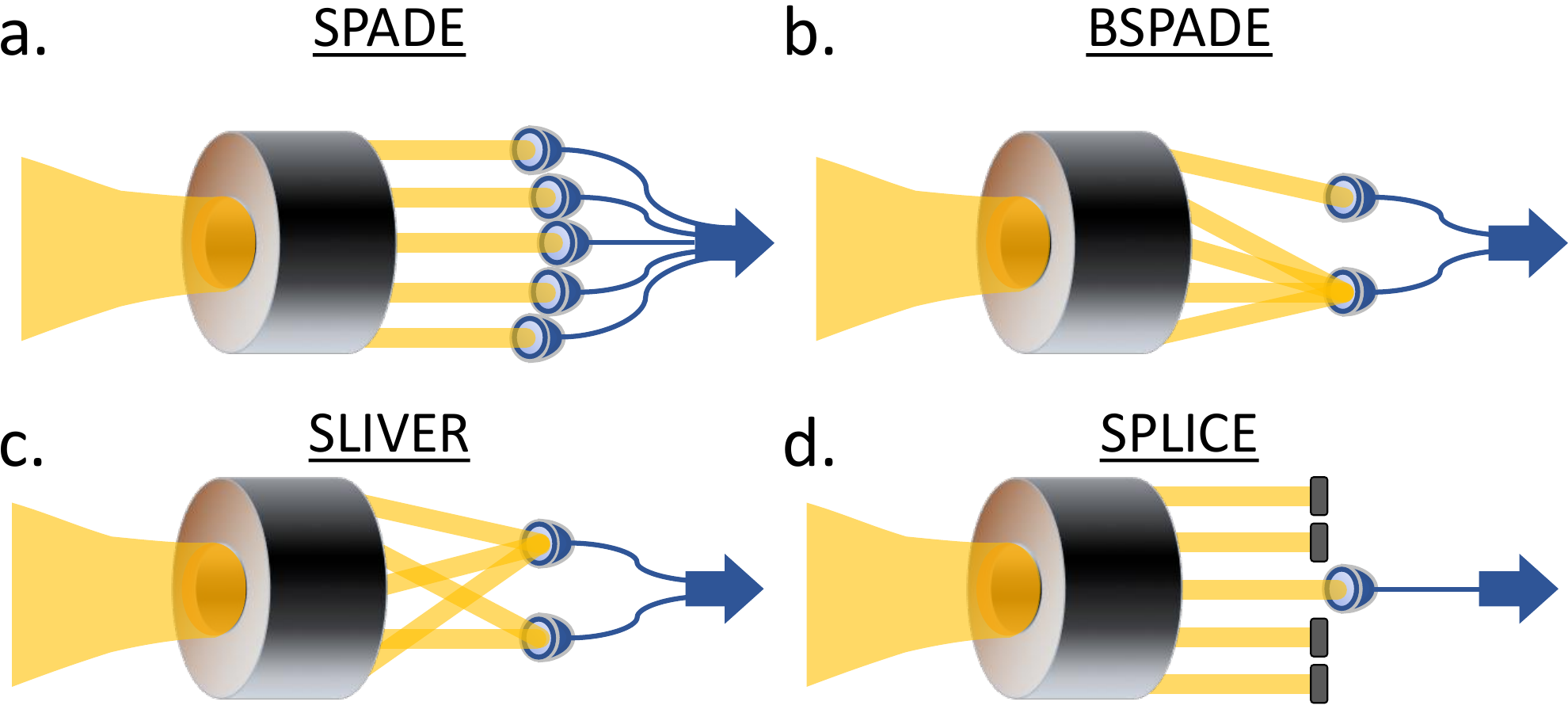}
	 \caption{Optical receiver paradigms for spatial mode processing. 
     }
    \label{fig:Receivers}
\end{figure}

Achieving passive sub-Rayleigh resolution for quantitative optical sensing requires a dedicated physical device that implements the required mode decomposition. This section discusses various structured receiver concepts and physical implementations of spatial mode processors (Fig.~\ref{fig:ModeSorters}). 

\subsection{Mode Processing Receiver Paradigms}
\teal{Measuring the input light field in a complete orthonormal spatial mode basis, as prescribed by theoretically ideal SPADE, is not only prohibitive in experimental practice, but also unnecessary for many practical tasks. It often suffices to separate and measure only some of the modes, while allowing other modes to remain mixed or even be lost (Fig.~\ref{fig:Receivers}).}
An example of a simplified receiver is binary SPADE (BSPADE), in which a single target mode is coupled to a dedicated detector while the entire orthogonal complement of the received field is collected on a bucket detector~\cite{Tsang16}. BSPADE implementations most often designate either the fundamental PSF mode or the first derivative mode as the target mode~\cite{Kerviche2017a}. A further simplification is offered by  super-resolved position localisation by inversion of coherence along an edge (SPLICE) \cite{Tham17}, which simply discards the component that is orthogonal to the mode of interest. Another two-detector receiver design, super localization by image inversion interferometry (SLIVER), achieves “separation of the image-plane field... into its symmetric and antisymmetric parts with respect to inversion in the image plane about the optical axis”~\cite{Nair2016b}. \teal{For the HG basis, this means separation of even and odd HG mode components.} The power of SLIVER for sub-Rayleigh sensing stems from the fact that while the bulk of the energy exists in the symmetric subspace of the optical field (e.g., even HG modes), the information about source locations, sizes and distances is primarily contained in the antisymmetric (odd) modes.

In Table~\ref{tab:relativeFIcomparison} we report the quantitative scaling laws for the Fisher information on the relative error for two-point-source separation estimation compared against Chernoff exponents for one-vs-two point source detection, to lowest order in $\theta\Delta k$ with a Gaussian PSF. Table~\ref{tab:relativeFIcomparison} extends the results of Eqs.~(\ref{eq:xi_Q})-(\ref{eq:FI_direct_rel}) to all of the receiver designs depicted in Fig.~\ref{fig:Receivers}. Noteworthy observations include the fact that direct imaging exhibits a quadratic scaling deficiency in terms of $(\theta\Delta k)$ compared to the quantum limit for both sensing tasks; SPADE is the only receiver of the four that saturates the quantum limit under all values of $\theta\Delta k$, as it implements complete basis decomposition of the optical field; the other three receivers approach the quantum limits in the deep sub-diffraction regime $\theta \Delta k\to 0$, when the relevant information is compressed into only the two lowest-order spatial modes.

\begin{table}[h!]
\setlength{\tabcolsep}{1pt}
	\centering
	\begin{tabular}{ |p{1.5cm}|p{3.7cm}|p{3.7cm}| } 
		\hline
		Method & FI on relative error & Chernoff exponent\\ 
		\hline\hline
        Quantum Limit & $\theta^2 \Delta k^2$~\cite{Tsang16} & $\theta^2 \Delta k^2/4$~\cite{lu2018quantum} \\ 
		\hline
        Direct \quad Imaging & $2\theta^4\Delta k^4+O(\theta^6\Delta k^6)$~\cite{yang2017} & $\theta^4\Delta k^4/16\,+$ $O(\theta^6\Delta k^6)$~\cite{lu2018quantum}\\ 
        \hline
		SPADE & $\theta^2 \Delta k^2$~\cite{Tsang16} & $\theta^2 \Delta k^2/4$~\cite{lu2018quantum}\\
        \hline
        BSPADE & $\theta^2\Delta k^2 +O(\theta^4\Delta k^4)$~\cite{Tsang16} & $\theta^2\Delta k^2/4+ O(\theta^4\Delta k^4)$\cite{lu2018quantum}\\
		\hline
		SLIVER & $\theta^2 \Delta k^2+O(\theta^4\Delta k^4)$~\cite{Nair2016b} & $\theta^2\Delta k^2/4+O(\theta^4\Delta k^4)$~\cite{lu2018quantum}\\
		\hline
		SPLICE & $2\theta^2 \Delta k^2/\pi+O(\theta^4\Delta k^4)$~\cite{Tham17} & $\theta^2\Delta k^2/2\pi\,+$ $O(\theta^4\Delta k^4)$ \\
		\hline
	\end{tabular}
\caption{Modified FI and CE for one-vs-two point source detection, to lowest order in $\theta\Delta k$ with a Gaussian PSF. 
All results are obtained from cited sources except the SPLICE CE, which we derived from~\eqref{eq:CE} using the mode projection probabilities from Ref.~\cite{Tham17}.}
\label{tab:relativeFIcomparison}
\end{table}

\subsection{Demultiplexing techniques}
\label{sec:SPADE_techniques}
In this subsection we review several of the most important methods for implementing SPADEs and SPADE-like devices that achieve the crucial pre-detection spatial-mode processing. We describe the technological development of each class of device, taking note of their suitability in terms of six desirable features for spatial-mode pre-processing: 
\begin{itemize}[noitemsep,nolistsep]
    \item low optical loss; 
    \item high sensitivity (ideally single-photon); 
    \item low modal crosstalk; 
    \item high mode depth (how many modes can a device sort at once?); 
    \item mode versatility (can the device be custom designed to sort a desired mode basis?); 
    \item mode reconfigurability (can a device be reconfigured to sort different mode bases?).
\end{itemize}

\begin{figure}[t]
\includegraphics[width=\columnwidth]{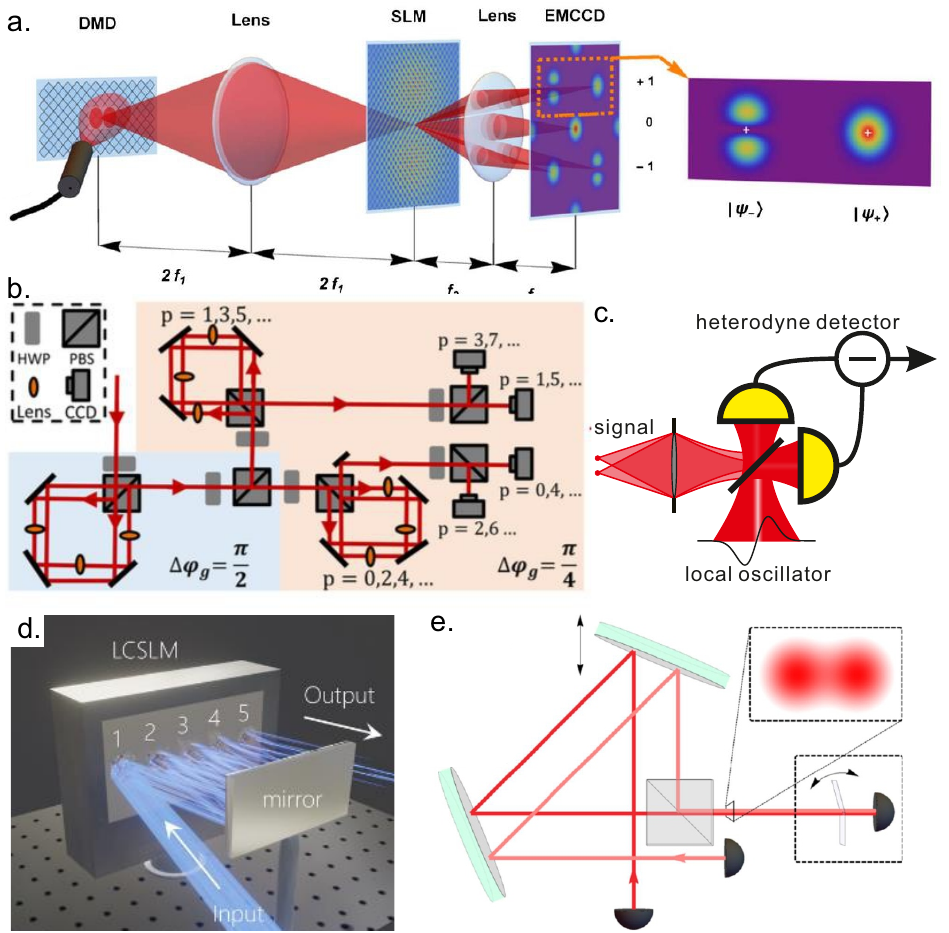}
	 \caption{Hardware implementations of spatial mode demultiplexers. (a) Holographic mode sorting of Hermite-Gauss modes~\cite{Paur16}. (b) Interferometric sorting of Laguerre-Gaussian mode by parity~\cite{Gu2018a}.  (c) Heterodyne detection of the mode of interest \cite{yang2016far} (d) Multi-plane light conversion~\cite{Kupiyanski2023}. (e) SPLICE projective measurement for point source separation estimation~\cite{Tham17}. \blue{Panel (b) reprinted with permission from~\cite{Gu2018a}; source: \url{https://doi.org/10.1103/PhysRevLett.120.103601}. Copyright 2018 by the American Physical Society. Panel (d) adapted from~\cite{Kupiyanski2023}; licensed under a Creative Commons Attribution (CC BY) license. Panel (e) reprinted with permission from~\cite{Tham17}; source: \url{https://doi.org/10.1103/PhysRevLett.118.070801}. Copyright 2017 by the American Physical Society.}}
     \label{fig:ModeSorters}
	 \end{figure}

\subsubsection{Interferometric Techniques}
\label{sec:Interferometry}
Spatial-mode demultiplexing was born out of pioneering work from the 1990s on HG-LG mode conversion~\cite{Beijersbergen1993a} and mode selection~\cite{Harris1994} in Gaussian laser beams. In 2002, Leach \emph{et al.} proposed a Mach-Zehnder interferometer (MZI) setup that could sort between even and odd orbital angular momentum (OAM) modes with (in principle) single-photon sensitivity~\cite{Leach2002a}. OAM constitutes one of the two degrees of freedom in a transverse optical field and can be mapped to the azimuthal index of the LG basis, so sorting these discrete, orthogonal modes can be used to transmit or detect information bearing signals. Furthermore, the MZI could be cascaded to sort higher-order parity and therefore increase the number of sorted modes. This cascading technique was extended to multi-mode, full-field sorting by converting between the radial (index $p$) and azimuthal (index $l$) domains via manipulation of the Gouy phase
of a propagating Gaussian beam~\cite{Zhou2017,Fu2018,Gu2018a}. Other groups pursued simpler MZI~\cite{Abouraddy2011} and Michelson interferometer~\cite{Malhotra2018} setups that traded experimental complexity for sensitivity, sorting an arbitrary number of modes but requiring many photons to build up an interferogram. 

Interferometry has most notably been applied for sub-Rayleigh imaging in the form of SLIVER [Fig.~\ref{fig:ModeSorters}(b)]. The image inversion is performed using a pair of lenses in either a MZI~\cite{Tang16} or a co-propagating configuration~\cite{Larson2019}. SLIVER saturates the quantum Cr\'amer-Rao bound as the separation between two point sources goes to zero, and modifications were identified to achieve the quantum limit for larger separations~\cite{Nair2016c} and for 2D imaging~\cite{Ang17}. 

\subsubsection{Mode Projection Techniques}
\label{sec:ModeProjection}
An alternative to simultaneously sorting the spatial modes of the optical field is to project just one mode to a detector. While this approach is inherently lossy, since all other modes of the field are discarded, focusing on just one mode can provide exceptional loss, sensitivity and crosstalk performance for that mode. The first example of mode projection for sub-Rayleigh imaging was that of Tham \emph{et al.}, which used a split phase plate to effectively map a signum function onto the PSF, thereby coupling only an antisymmetric component of the incident field to the single-pixel detector~\cite{Tham17} [Fig.~\ref{fig:ModeSorters}(e)]. This approach was termed super-resolved position localization by inversion of coherence along an edge (SPLICE). In another experiment, Bouchard \emph{et al.} achieved >99\% fidelity for full azimuthal and radial LG mode projection using a phase and amplitude hologram~\cite{Bouchard2018}. Switching the target mode throughout the observation can accrue additional information about the scene. This strategy was employed in an adapted version of SPLICE~\cite{bonsma-fisher_realistic_2019}. 

An alternative method for adapted mode projection is to use a balanced heterodyne detector with the local oscillator prepared in the mode of interest. The output photocurrent of such a detector is proportional to the amplitude of the signal mode matching that of the local oscillator [Fig.~\ref{fig:ModeSorters}(c)]. The advantage of this technique is that the local oscillator is relatively easy to prepare in an arbitrary spatial mode using a spatial light modulator (SLM) or a Fabry-Perot cavity. A shortcoming is that, like all mode-projection techniques, the method permits detecting only one mode at a time. This problem can be addressed by means of digital holography, i.e.~using a spatially broad local oscillator and acquiring the amplitude and phase of the interference pattern at each point of the detection plane. The content of any arbitrary spatial mode can then be computed by digital processing. More seriously, though, heterodyne detection adds shot noise of the local oscillator to the detection photocurrent. This leads to loss of advantage in terms of per-photon Fisher information with respect to direct imaging, with the exception of rare cases of very powerful signal sources~\cite{yang2017}. However, this approach could still be of benefit if the main cause of the resolution loss is caused by classical detection noise.  

\subsubsection{Holographic Techniques}
\label{sec:Holography}
A more sophisticated approach to mode demultiplexing involves holography [Fig.~\ref{fig:ModeSorters}(a)]. The first method for holographic mode sorting used forked holograms to flatten the wavefront of only certain OAM modes, allowing select modes to be coupled into a single mode fiber with a focusing lens~\cite{Mair2001a}. Gibson \emph{et al.} multiplexed a set of forked computer generated holograms (CHGs) to perform OAM  sorting on eight modes~\cite{Gibson2004a}. A downside to this type of planar hologram is an inverse scaling of the photon efficiency with mode number. This problem was resolved by the introduction of the Cartesian-to-log-polar transformation, which unwraps and phase-flattens the OAM modes into plane waves with different $k$ vectors~\cite{Berkhout2010a}. This efficient multiplexing allowed for experimental demonstrations of sorting up to 50 modes with minor loss of channel capacity~\cite{Lavery2012a}. In addition to improvements in loss and mode depth, the modal crosstalk was reduced to below 10\% by incorporating a holographic fanout that effectively increases the initial mode size to consequently reduce the far-field spot size that is to be coupled out for detection, minimizing the far-field mode overlap with neighboring modes~\cite{OSullivan2012a,Mirhosseini2013a}. Following from this development, a simple three-mode CHG was used in the first experimental SPADE demonstration to beat the mean squared error of direct imaging by >20x in the sub-Rayleigh regime, highlighting the advantages of holography for sensitivity and crosstalk~\cite{paur2016achieving}.

\subsubsection{Multiplane Techniques}
\label{sec:MPLC}
Unlike two dimensional plate holograms or CHGs, three dimensional volume holograms continuously process the optical wavefront as it propagates, potentially enabling greater multiplexing capabilities with just phase modulation~\cite{Wang2018}. A clever way to access this benefit while avoiding the design and fabrication challenges associated with volume holograms is to use a sequence of phase-only CHGs that are each separated by a fixed distance [Fig.~\ref{fig:ModeSorters}(d)]. In 2010, Morizur \emph{et al.} showed that alternating these two building blocks (phase masks and free-space propagation) just three times enabled the full phase and amplitude processing required for SPADE with $\sim10\%$ crosstalk~\cite{Morizur:10}. In principle, an infinite number of alternative phase masks and free-space offsets is proven to enable universal spatial-mode conversion~\cite{Morizur:10,Borevich1981}. This strategy, now called multi-plane light conversion (MPLC), has become one of the most promising SPADE implementations~\cite{Labroille:14,fontaine2019,boucher_spatial_2020,Tan2023}, having reached mode counts of 1035~\cite{fontaine2020laguerre} and reduced crosstalk to <1\% when sorting 10-16 modes~\cite{Boucher2021}. 
MPLC can be designed either as a high-precision, fixed-basis device~\cite{Boucher2021} or as a reconfigurable mode sorter that dynamically updates the mode basis by refreshing the phase masks displayed on a spatial light modulator (SLM)~\cite{Kupiyanski2023,Ozer2024}. \blue{Moreover, the MPLC can be trained with the direct objective of optimizing the CRB --- so the quantum optimal basis is not pre-determined but is unveiled in the process of training. This approach has proven effective in point-source separation estimation \cite{filipovich2026learned} and image reconstruction \cite{warke26}.}

\subsubsection{Other approaches}
A few other techniques have been considered for SPADE-like resolution enhancement in the sub-Rayleigh regime. A truly quantum method was proposed in Parniak~\emph{et al.}, where photon pairs were experimentally detected for Hong-Ou-Mandel coincidences resulting from spatially-mediated variations in photon distinguishability, though this approach would require quantum nondemolition measurements to achieve superresolution with thermal sources~\cite{parniak2018beating}. A nonlinear optics approach was demonstrated that used mode-selective sum-frequency generation to preferentially upconvert photons under one sub-Rayleigh hypothesis over the other~\cite{Zhang2020c}. Another well-established linear optical device that could be used as a spatial-mode sorter is the photonic lantern~\cite{salit2020experimental,eikenberry2024photonic,kim2025sky}. Finally, there is some evidence that the performance of direct imaging can be partially restored back toward the quantum limit without any spatial-mode sorting, instead using either PSF engineering prior to the detection~\cite{Paur2018b} or post-processing only the signal that arrives near zero points in the PSF~\cite{Paur2019}. 

\section{Experimental achievements} \label{sec:O4}
Most of the theoretical ideas presented in Sec.~\ref{sec:O2} have been successfully tried experimentally using the tools described in Sec.~\ref{sec:O3}.
\subsection{Point source separation}
Immediately after the publication of the preprint of the Tsang {\it et al.}~idea \cite{Tsang16} in November 2015, four research groups implemented proof of principle  experiments in various settings modeling a subwavelength imaging system \cite{Tang16,paur2016achieving,Yang16,Tham17}. These papers differed from each other in experimental techniques, including light source preparation and detection, as well as the degree of rigor in evaluating the FI. In Tang {\it et al.} \cite{Tang16} and Tham {\it et al.}~\cite{Tham17}, the two sources were modelled by the two parallel Gaussian beams 
of orthogonal polarizations, which emulated two incoherent sources. In Pa\'{u}r {\it et al.} \cite{paur2016achieving}, two point sources of variable separation were produced by a digital micromirror device (DMD)  illuminated by a He-Ne laser. Incoherence was simulated by the DMD displaying one source at a time and switching between the two with a high frequency. Yang {\it et al.}~used commercial double slits for educational labs, exploring both the coherent and incoherent regimes \cite{Yang16}. For mode projection, the four experiments used the full range of techniques described above: image inversion interferometry \cite{Tang16}, holographic mode separation \cite{Paur16},  heterodyne detection \cite{yang2016far} and SPLICE \cite{Tham17}. Pa\'{u}r {\it et al.}~\cite{paur2016achieving} and Tham {\it et al.}~\cite{Tham17} implemented quantum detection, which enabled evaluating the per-photon Fisher information and thus rigorously demonstrating advantage over DI. 

A further interesting contribution of a Tham {\it et al.}~\cite{Tham17} was to reconcile the theoretical prediction of the uncertainty for DI tending to infinity according to the Cram\'er-Rao bound (Fig.~\ref{QCRB}) and the experimental (and common sense) observation that this uncertainty does not exceed the PSF width. The explanation offered is that ``the $\theta$ and $-\theta$ cases are physically indistinguishable; therefore, what is really being estimated is the absolute value $|\theta|$'', which cannot be negative. As a result, the estimation is inherently biased, while the  Cram\'er-Rao bound applies to unbiased estimates only (\emph{cf.}~Sec.~\ref{sec:O2}.\ref{sec:biased}).



The problem of point-source separation measurement has been revisited recently by Boucher {\it et al.}~\cite{boucher_spatial_2020}, who used a commercial MPLC mode sorter (Cailabs) to direct each HG$_{mn}$ mode up to $m,n\le 2$ into single-mode fibers. This  allowed measuring multiple  Hermite-Gaussian modes simultaneously and extend the sensitivity of the method outside the range of very small displacements. Rouvi\`ere {\it et al.}~\cite{rouviere2024ultra} refined these measurements further, achieving a precision of 20 $\mu$m with a 1-mm PSF and 3500 incoming photons. In an intense beam regime ($10^{13}$ incoming photons), they demonstrate a 20 nm sensitivity, about five order of magnitude better than the diffraction limit. 

Wadood \emph{et al.}~performed an experiment demonstrating the
estimation of the separation between partially coherent sources in the
sub-Rayleigh regime (cf.~Sec.~\ref{sec:O2}.\ref{sec:PartCoh}) \cite{wadood21}. Their theoretical study and
experimental condition generalize prior works by assuming that both
the separation and the brightness of each source are unknown. The
partially coherent sources are simulated by a laser beam and a
SLM. For both incoherent sources and partially coherent sources with a
negative degree of coherence $\gamma = -0.75$, the reported
experimental errors are comparable with the quantum
limits. Remarkably, the errors for the negatively correlated sources
are below those for incoherent sources, in agreement with theoretical
predictions \cite{wadood21,tsang_comment19,larson19}.

Donohue {\it et al.} extended SPADE to the time domain, resolving  temporal and spectral separations between incoherent pulses at the single-photon level \cite{donohue2018quantum}. Separations as small as one-tenth of the pulses’ optical bandwidth were measured by projecting the pulses onto the HG$_{01}$ mode in the time-frequency domain  using mode-selective sum-frequency generation with a spectrally shaped reference pulse in a nonlinear waveguide. The component of the signal mode that matched the reference experienced up-conversion, so the energy of the up-converted signal determined the intensity of that component.  The same group later extended their study towards multiple parameters of the signal pulse, demonstrating simultaneous estimation of the timing centroid, offset, and intensity imbalance of the two incoherent pulses \cite{ansari2021achieving}. De \emph{et
  al.}\ extended these experiments to partial coherence settings \cite{de21}. 

All of the experimental demonstrations we have reviewed in this section utilize laser light to emulate incoherent sources, either by using multiple lasers or by breaking the temporal coherence of a single laser beam via speckle reduction on a rotating defuser, sequential detection of multiple parts of the scene, or a related method. An important exception is Mitchell {\it et al.} \cite{mitchell2026quantum}, who performed a SLIVER-like measurement on pairs of real 40 nm fluorescent beads using an optical microscope equipped with a deformable mirror for adaptive optics. The authors reported >10x enhanced Fisher information content in their measurement data compared with direct imaging for samples down to a separation of 5 nm. Interestingly, they note that this advantage only appears when the emission path is filtered to collect only azimuthally polarized light, owing to the effect of fluorescent dipoles on optical interference. This work helps clarify the pathway toward useful, passive sub-diffraction microscopy based on spatial mode processing.

\subsection{Parameter estimation for more complex scenes}

Santamaria \emph{et al.}~\cite{santamaria2024} investigated the estimation of the separation and relative intensity of two incoherent point sources with unequal brightness in the single-photon regime, using an MPLC by Cailabs(cf.~Sec.~\ref{sec:O2}.\ref{sec:PowerImbalanced}). By aligning the optical system to the brighter source so that it couples to the $HG_{00}$ mode, and measuring only the $HG_{01}$ and $HG_{10}$ modes, they demonstrate sub-Rayleigh precision in the unbalanced regime. Experimentally, they recorded the total number of counts as a function of the beam separation at fixed power imbalance and observe the expected quadratic dependence in the sub-Rayleigh limit. Conversely, by varying the relative intensity at fixed separation, they obtain a linear response when the imbalance is large.



Deshler {\em et al.} reported the first experimental demonstration of a coronagraph design that reaches the quantum limits of detecting and localizing exoplanets close to a bright star at angular separations below the Rayleigh limit~\cite{Deshler2025_Expt}. \teal{The MPLC first sorts out the fundamental mode  containing most of the emission energy from the star. That mode is then blocked and the light is then reflected back through the MPLC \blue{into a separate output port} to undo the mode sorting transformation~\cite{Deshler2025_Expt}. The resulting optical field, when detected on a camera, conveys an image of the scene with the starlight largely eliminated. The experimental system successfully localizes an artificial exoplanet at sub-diffraction separations with a 1000:1 star–planet contrast. This approach surpasses several well-known coronagraph strategies such as the Lyot stop, Vortex and PIAACMC, and demonstrated performance approaching fundamental quantum limits.}


Several experiments have demonstrated the feasibility of approaching the quantum precision limits for three-dimensional localization predicted in Sec.~\ref{sec:O2}.\ref{sec:3Dloc}. Zhou \emph{et al.}~\cite{zhou2019quantum} implemented a radial mode sorter to resolve axial separations between two incoherent sources, using LG modes to extract depth information. The scheme achieved quantum-limited precision for vanishingly small separations without sacrificing photon efficiency. Independently, \v{R}eh\'a\v{c}ek \emph{et al.}~\cite{rehacek2019optimal} showed that defocused intensity measurements with a conventional camera, combined with maximum-likelihood estimation, can also saturate the QCRB for axial localization. Their experiment achieved depth resolution exceeding the classical diffraction limit by over three orders of magnitude. 

Complementing these approaches, Hu \emph{et al.}~\cite{hu2023experimental} performed full 3D localization of a single emitter. \blue{They derived the 3D QFI matrix for emitters whose point spread function is a single LG mode or a rotating superposition of such modes --- a class of fields introduced by Schechner, Piestun and Shamir~\cite{schechner1996wave,piestun2000propagation}  --- and established how much of that information is accessible to ordinary intensity detection. In a proof-of-principle experiment with modes generated on a spatial light modulator and recorded on an ICCD camera, a maximum-likelihood estimator incorporating a realistic detector-noise model was shown to attain the corresponding Cram\'er-Rao bound for} 
all three spatial coordinates. For single LG mode PSFs, their technique achieved a two-fold improvement in transverse ($x$, $y$) precision and up to a twenty-fold enhancement in axial ($z$) localization compared to the Gaussian PSF baseline, in good agreement with Eq.~(\ref{eq:QFI3D}), even in the presence of realistic optical aberrations. \blue{The rotating superposition eliminated the $\pm z$ ambiguity and improved axial precision near focus to some extent, but intensity detection recovered only a small fraction of its 3D QFI.}

Tan \emph{et al.} \cite{Tan2023} studied the localization of two incoherent optical point sources in two dimensions (Sec.~\ref{sec:O1}.\ref{sec:MultiLoc}). In this regime, when the distance between the two sources is within the diffraction limit, demultiplexing on a Hermite-Gaussian basis induces ambiguity due to the symmetry of the system. However, by taking advantage of the natural small asymmetry of the MPLC (Cailabs), they were able to resolve the full 2D positions of both sources rather than only their separation. The experimental results were emulated from a calibration of the apparatus using a single source spanning the 2D field of view, and provide position estimations for equal-power sources.

\subsection{SPADE for hypothesis testing}

A few additional works have utilized experimental mode-sorting setups to demonstrate sub-diffraction hypothesis testing between one and two point sources (Sec.~\ref{sec:O1}.\ref{sec:1vs2discrimination}). Zhang {\it et al.} demonstrated a novel scene classifier based on sum-frequency generation (SFG) \cite{Zhang2020c}. By spatially shaping an external pump, the classifier can selectively optimize the sum-frequency efficiency for one hypothesis or the other. Spectral filtering the sum-frequency output 
and counting the upconverted photons provides mode-selective information about the scene, achieving a desired discrimination accuracy with >100x fewer photons than ideal direct imaging.

Zanforlin {\it et al.}~\cite{zanforlin2022optical} achieved near-quantum-limited asymmetric hypothesis testing with a two-detector spatial demultiplexing measurement. The work implemented a receiver proposed by Lupo {\it et al.}~\cite{Lupo2020} for interferometric projection onto orthogonal two-mode basis states (e.g., an even and an odd mode, reminiscent of SLIVER), which is potentially easier to build than a multimode SPADE or SLIVER. They confirmed the predicted quadratic scaling advantage in relative entropy over direct imaging down to a source brightness ratio of ~1\%, beyond which the receiver attained a constant-factor advantage in discrimination accuracy. 

Wadood {\em et al.} showed an improvement over idealized direct imaging in average error probability for symmetric hypothesis testing of one vs. two point sources~\cite{wadood2024experimental}. This work demonstrated the robustness of mode-sorting measurements in the presence of modal crosstalk, which the authors quantified at 0.8\% in their experiment.


\subsection{SPADE for imaging}
Prior to discussing the application of SPADE to enhance imaging resolution, a rigorous definition of the latter must be introduced. Consider first a coherent object with the field amplitude $E(\vec r)$ viewed with a microscopic objective with numerical aperture NA. The objective's aperture is located in the Fourier plane of the object and modifies the Fourier image $\tilde E(\vec k_\perp)$ of the object by multiplying it by the amplitude transfer function of the top-hat shape $\mathrm{ATF}(\vec k_\perp)=\theta\left(\frac{2\pi}\lambda\mathrm{NA}-|k_\perp|\right)$. The resulting image, obtained via inverse Fourier transform, is then a convolution 
$E'(\vec r)=E(\vec r)*\mathrm{APSF}(\vec r)$, with the amplitude point-spread function $\mathrm{APSF}(\vec r)$ being the inverse Fourier image of $\mathrm{ATF}(\vec k_\perp)$ --- an Airy disk.

If the object is incoherent and defined by the intensity function $I(\vec r)$, the image is the convolution $I'(\vec r)=I(\vec r)*\mathrm{PSF}(\vec r)$, where $\mathrm{PSF}(\vec r)=|\mathrm{APSF}(\vec r)|^2$ is the point-spread function. In the Fourier domain, this can be rewritten as 
\begin{equation}\label{Iprime}
    \tilde I'(\vec k_\perp)=I(\vec k_\perp)\mathrm{OTF}(\vec k_\perp),
\end{equation}
where 
\begin{equation}\label{OTF}
    \mathrm{OTF}(\vec k_\perp)=\mathcal F[\mathrm{PSF}(\vec r)]=\mathrm{ATF}(\vec k_\perp)*\mathrm{ATF}(\vec k_\perp)
\end{equation}
is the optical transfer function: the convolution of the top-hat amplitude point-spread function with itself. This is a function of a nearly conical shape with the support $k_\perp\le\frac{4\pi}\lambda\mathrm{NA}$. 

This result is fundamental in that no measurement, including SPADE, can recover Fourier components of the object intensity that are above the cut-off $k_c=\frac{4\pi}\lambda\mathrm{NA}$ in a linear, passive, near-field setting. However, simple direct imaging does not make optimal use even of those Fourier components that are transmitted through the objective: multiplication (\ref{Iprime}) with the conical OTF (\ref{OTF}) suppresses high Fourier components, leading to further loss of resolution. In the low-noise limit, this additional loss can be recovered by numerical processing of the image, such as deconvolution. On the other hand, when the image is contaminated by classical or quantum noise, an optimized measurement --- such as SPADE --- can help reduce the effect of this noise and hence improve the resolution. However, the advantage that can be gained in this manner cannot exceed the resolution obtainable in the absence of noise by means of deconvolution. 

\teal{The above fundamental limitation on the acquisition of Fourier components may appear to be in disagreement with the argument of Sec.~\ref{sec:O2}.\ref{sec:moments} that SPADE allows determination of arbitrary moments of the source intensity distribution, and hence the distribution itself, with arbitrarily high precision. However, accurate reconstruction of a distribution form its moments requires that the distribution has a finite support --- i.e.~that the object imaged is of \emph{finite size}. Such an object can be expressed as a product of an infinitely wide distribution with a top-hat function, and its Fourier image --- as a convolution of the Fourier image of that distribution with a sinc function. As a result, high-frequency Fourier components of the object are admixed into low spatial frequencies and may leak through the aperture. This leakage, albeit weak, provides the observer with information about these components. }



\begin{figure}[t]
	 \begin{center}
	 	\includegraphics[width=\columnwidth]{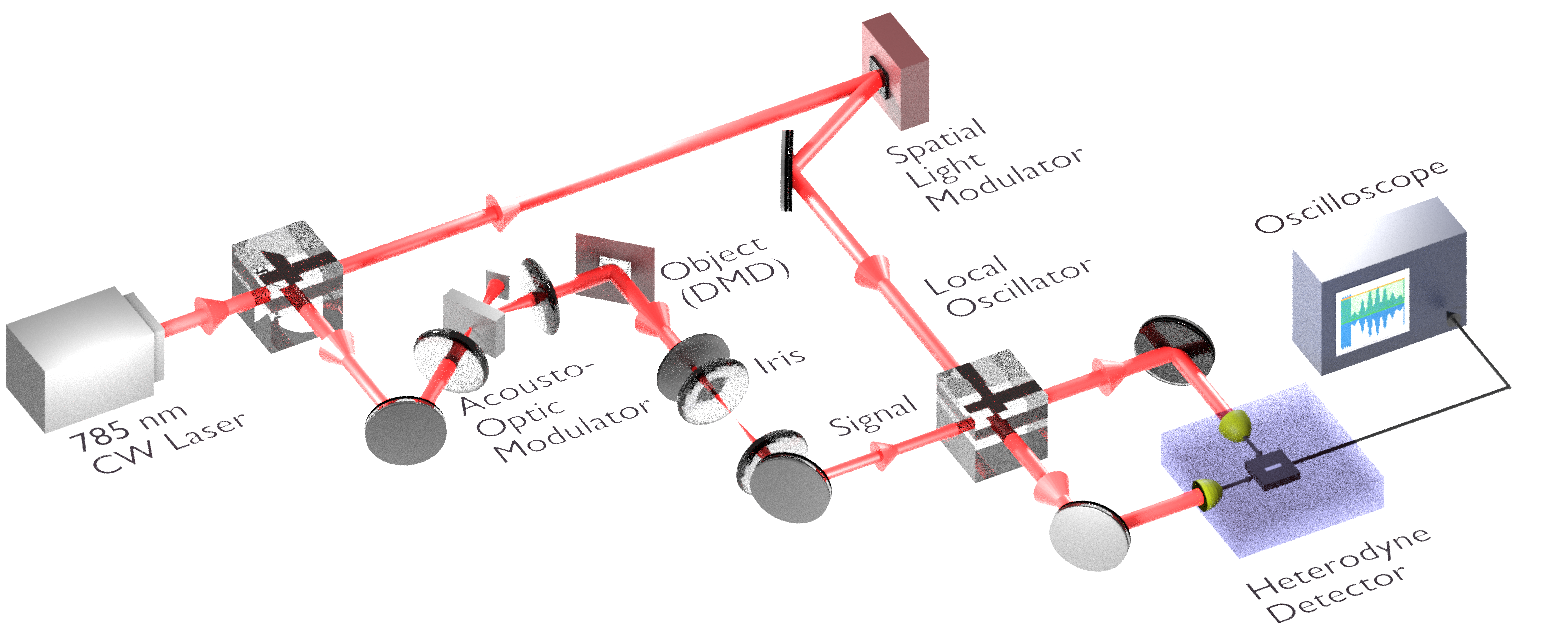}
	 	\includegraphics[width=\columnwidth]{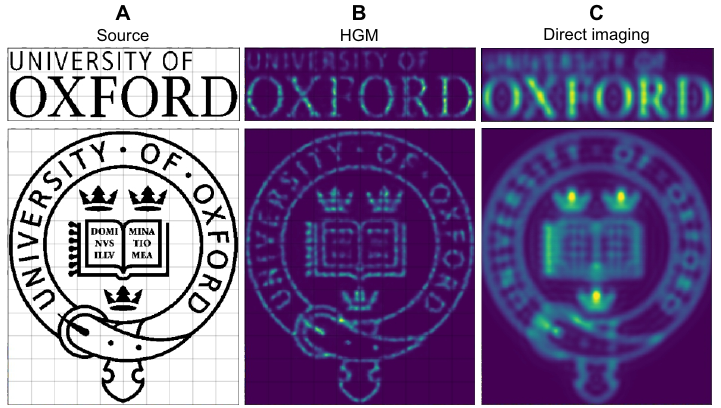}	     
	 \end{center}
	 \caption{\label{fig:OxfordExptFig} The experiment of Pushkina {\it et al.}. Top: Setup. Bottom: Results --- the original (A), SPADE-reconstructed (B) and camera (C) images of the Oxford logo  test sets. \blue{Adapted from \cite{pushkina2021superresolution}. Copyright 2021 American Physical Society.}}
	 \end{figure}

Experimentally,  proof-of-concept image reconstruction was realized  in a model microscopic setting (Fig.~\ref{fig:OxfordExptFig}), achieving a resolution enhancement by a factor of two compared to the diffraction limit \cite{pushkina2021superresolution}.
The ``object'' was the University of Oxford logo divided into multiple frames (shown by the gray grid), each of which was displayed sequentially on a digital micromirror device illuminated by a laser. The diffraction limit was imposed by a narrow iris diaphragm in the reflected field. Balanced heterodyne detection was used to measure the amplitude and phase of Hermite-Gaussian components of the reflected light. The 441 local oscillator modes, ranging from HG$_{0,0}$ to HG$_{20,20}$, were prepared sequentially by means of a spatial light modulator. For image reconstruction, a deep feedforward neural network was designed that took the complex heterodyne output photocurrents as inputs and produced the image as output. The neural network was trained using various semi-random patterns (rectangles, lines, ellipses, etc.) displayed on the DMD and the corresponding heterodyne output photocurrent sets.

This work was followed by a demonstration of SPADE imaging of incoherent objects \cite{frank2023passive}, with the observed resolution enhancement by $\sim\times3$. The experiment followed a similar scheme, but the overcomplete basis of Ref.~\cite{Tsang17} was used for reconstruction. Incoherent objects were simulated by displaying pixels on a DMD one-by-one and measuring the heterodyne photocurrent powers associated with various HG modes and their superpositions. To simulate a particular pattern from the training and test sets, the powers associated with individual pixels that formed the pattern have been added. 

A follow-up experiment \cite{duplinskiy2024tsang} by the same group studied compatibility of SPADE with another resolution enhancement technique: 
image scanning microscopy (ISM) \cite{sheppard1988super}. In this method, the sample is scanned under a focused illumination beam. Snapshots recorded at each
scanning step are combined into the final image using the so-called pixel reassignment routine. Because only the light within the narrow illumination spotlight is detected in each snapshot, the resolution of the resulting image is improved compared to widefield illumination.

The study by  Duplinskiy {\it et al.}~\cite{duplinskiy2024tsang}   reconstructed each snapshot using SPADE and then used pixel reassignment to combine them. The resolution enhancements due to focused illumination and detection via SPADE are then cumulative, because the effective image reconstruction PSF is the product of the illumination and detection PSFs. However, in this simulation of a \emph{bona fide} microscope, the SPADE PSF was much narrower than the illumination PSF, so the cumulative resolution enhancement effect in comparison to SPADE with widefield illumination was limited by $\sim10\%$. On the other hand, the reconstructed image exhibited significantly higher quality and lower noise. Application of standard deconvolution methods to that image further improved the resolution to about a factor of 2 compared to widefield SPADE.

The above experiments were performed in a macroscopic setting simulating a microscope, and did not show any comparison with quantum benchmarks. Very recently, however, measurements of high spatial frequency components of a microscopic image with verifiable per-photon quantum advantage with respect to DI have been demonstrated \cite{gong2026passive}. In that work, a custom phase plate separated the Fourier space of the object into separate regions, each of which was imaged separately. In this way, detection of high spatial frequency components is not contaminated by shot noise from low Fourier components, resulting in enhancement of the corresponding FI, albeit at the expense of the FI for lower frequency components.

Finally, a recent imaging result was reported by Kim {\em et al.}, who utilized a photonic lantern (PL) to perform simultaneous imaging and spectroscopy of a \emph{bona fide} stellar accretion disk~\cite{kim2025sky}. The experiment relied on the mode-mixing capability of a mode-selective PL to demultiplex the incident light into orthogonal output modes. Aided by high-order adaptive optics and time-domain post-processing to reduce the effects of atmospheric turbulence, 10 minutes of data from the PL-based SPADE measurement was used to reconstruct the an image of the disk and spatially resolve its rotation velocity field. This work confirms the utility of SPADE receivers for quantitative on-sky observation.

\section{Perspectives of the domain} \label{sec:O5}
\purple{[SG to add references.]} Over the past decade, passive super-resolution sensing using spatial mode sorters and related techniques has evolved from a primarily theoretical construct to a mature experimental paradigm with demonstrated performance approaching the quantum limits. What began with single-parameter estimation of point-source separation has expanded into a broad family of methods of analyzing diffraction-limited visual information, ranging from estimating object parameters and object classification according to sub-Rayleigh features to complete image reconstruction with sub-Rayleigh resolution. 

Looking forward, we envision further expansion of this approach into a broader framework of quantum-optimal design for sensing visual information: constructing measurements on the incoming electromagnetic field that access \emph{arbitrary} parameters of interest, or functions thereof, with maximal precision allowed by quantum physics. This perspective re-frames super-resolution into a structured estimation problem, closely aligned with developments in compressed sensing, Bayesian inference, and quantum multiparameter estimation theory.

While most of the estimation and hypothesis testing tasks considered in this review could be addressed using standard measurement bases, such as Hermite-Gaussian, the above framework requires custom-designed measurements and hence bespoke bases that are specifically constructed for the task at hand and the prior information available, such that the information of interest is concentrated in a small number of basis elements. This brings the mode sorter design to the fore. Multi-plane light conversion (MPLC) devices can be interpreted as \emph{diffractive optical neural networks}, where each phase plane represents a weight matrix implementing a linear transformation on the optical field \cite{lin2018all,hu2024diffractive,warke26}. Accordingly, machine learning methods can be used to optimize (\emph{train}) the MPLC surfaces in a task-specific way to extract maximum allowed information from every photon. When combined with electronic or optical non-linearities, such architectures may enable hybrid optical–digital inference engines that jointly optimize the measurement and estimation stages. 

This gives rise to  a new paradigm: \emph{quantum computational machine vision (QCMV)}, where artificial intelligence is incorporated into the physics of measurement and trained to maximize the precision for that measurement in application to the task at hand. QCMV can be beneficial for visual feature learning  under the conditions of scarce light --- for example, detecting a road sign at night, analyzing delicate biological samples under phototoxic limits, or discerning distant celestial bodies --- when the discrete nature of photons introduces quantum noise that constrains traditional imaging \cite{filipovich2026learned}. 

One can draw parallels comparing QCMV with quantum computing and quantum machine learning in that both involve complex processing of quantum states. An important difference is that a quantum computing task normally has a classical input, e.g.~a number that needs to be factorized in the case of Shor's algorithm. In QCMV, on the other hand, the input data are inherently quantum, and necessarily requires measurement. As a result, quantum advantage in vision does not primarily originate from quantum state preparation or quantum processing of classical data, but from optimizing the measurement performed on the optical field. Such advantage --- and practical benefits thereof --- may therefore be easier to achieve than in quantum computing, representing a ``low hanging fruit'' in quantum technology that was largely overlooked so far \cite{khan2025}. 


A further advantage of QCMV is the possibility of integration of \emph{adaptive estimation and turbulence compensation} \cite{Billaud-Cailabs} within the mode-sorting architecture itself, without relying on conventional wavefront correction. In this view, turbulence estimation, aberration mitigation, and super-resolution become parts of a single adaptive measurement loop, where Bayesian updates or real-time optimization determine the most informative measurement basis at each stage. Such strategies naturally connect to quantum measurement theory, where adaptive measurements are known to be essential for saturating multiparameter quantum Cramér–Rao bounds.  

\blue{One task in which QCMV has proven useful is that of quantum-optimal image estimation \cite{warke26}. While addressed theoretically for incoherent two-dimensional objects \cite{warke26,deshler26}, this  semiparametric quantum estimation~\cite{tsang19a,TsangAlbarelliDatta20} problem still needs to be generalized to three-dimensional settings and beyond the low-contrast approximation, as well as extended to a setting in which not only detection but also illumination is trainable. Experimental implementation of these approaches will likely bring about a new generation of microscopes and telescopes capable of precise estimation of high spatial frequency components in the photon-starved regime.} 

Alternative physical routes to super-resolution also remain an active area of exploration. \emph{Superoscillation} provides a conceptually distinct mechanism for engineering arbitrarily narrow focal spots and tailored point-spread functions, albeit with trade-offs in energy concentration and sidelobe structure. While superoscillatory fields do not evade quantum limits, their integration with mode sorting and estimation theory may offer new degrees of freedom in practical system design, particularly for applications where \emph{a priori} information about the sample --- such as its limited size --- is available \cite{tsang2022,chang2026super}.

Multi-aperture and distributed sensing architectures represent another promising direction. Local mode sorting at each aperture, combined with classical or quantum correlations across apertures, opens the door to entanglement-assisted sensing protocols that can outperform separable strategies in information extraction, with implications for astronomy, synthetic aperture imaging, and networked quantum sensors \cite{padilla2026superresolution}.


Finally, the work covered in this review assumes classical (thermal) nature of the light sources. While this is valid for remote sources such as stars, microscopic emitters, such as flourophore molecules or color centers, often emit fields with nonclassical (antibunched) properties. These properties can serve as signatures of individual emitters, enabling sub-Rayleigh resolution through photon statistics measurements \cite{gatto2014beating,tenne2019super}. This approach provides a complementary measurement paradigm, emphasizing that mode sorting is just one instance of a broader class of structured optical measurements.

\begin{backmatter}
    \bmsection{Funding} 
    We acknowledge financial support by the UK Research and Innovation (UKRI) under BBSRC Grant No.~BB/X004317/1 and EPSRC Grant No.~EP/X010929/1. MRG and SG acknowledge financial support by the Defense Advanced Research Projects Agency (DARPA) over a number of years, on the InPho, REVEAL, EXTREME, IAMBIC and QUINTISSA programs.

\bmsection{Disclosures} The authors declare no conflicts of interest.

\bmsection{Data Availability Statement} Data underlying the results presented in this paper are not publicly available at this time but may be obtained from the authors upon reasonable request.


\end{backmatter}


\bibliography{sample}

\end{document}